\documentclass[useAMS,usenatbib]{mn2e}
\usepackage{graphicx,epsfig,amssymb,amsmath,layout,verbatim,rotating,calc,mathrs
fs,natbib}
\usepackage{graphics}
\usepackage{rotate}
\usepackage{txfonts}

\newcommand{\ksk}{km~s$^{-1}$~kpc$^{-1}$ }

\newcommand{\ej}{E$_{J}$ }

\title[NGC1300 orbital analysis]{NGC~1300 Dynamics: III. Orbital analysis}
\author[P.A.~Patsis et al.]
{P.A.~Patsis,$^{1,2,3}$\thanks{patsis@academyofathens.gr (PAP); ckalapot@phys.uoa.gr
  (CK); pgrosbol@eso.org (PG)}
C.~Kalapotharakos,$^{1}$ and P.~Grosb{\o}l$^{3}$\footnotemark[1]\thanks{Based on
 observations collected at the European Southern
Observatory, Chile: program: ESO 69.A-0021.}\\
$^1$Research Center for Astronomy, Academy of Athens, Soranou Efessiou
    4, GR-115 27, Athens, Greece\\
$^2$ Observatoire Astronomique de Strasbourg, 11 rue de l'Universit\'{e}, 67000
Strasbourg, France\\
$^3$ European Southern Observatory, Karl-Schwarzschild-Str. 2, 85748 Garching,
Germany }

\date{Accepted ..........Received .............;in original form ..........}

\begin{document}

\maketitle

\label{firstpage}

\begin{abstract}
We present the orbital analysis of four response models, that succeed in reproducing morphological features of NGC~1300. Two of them assume a planar (2D) geometry with $\Omega_p$=22 and 16~\ksk respectively. The two others assume a cylindrical (thick) disc and rotate with the same pattern speeds as the 2D models. These response models reproduce most successfully main morphological features of NGC~1300 among a large number of models, as became evident in a previous study. Our main result is the discovery of three new dynamical mechanisms that can support structures in a barred-spiral grand design system. These mechanisms are presented in characteristic cases, where these dynamical phenomena take place. They refer firstly to the support of a strong bar, of ansae type, almost solely by chaotic orbits, then to the support of spirals by chaotic orbits that for a certain number of pattern revolutions follow an n:1 (n=7,8) morphology, and finally to the support of spiral arms by a combination of orbits trapped around L$_{4,5}$ and sticky chaotic orbits with the same Jacobi constant. We have encountered these dynamical phenomena in a large fraction of the cases we studied as we varied the parameters of our general models, without forcing in some way their appearance. This suggests that they could be responsible for the observed morphologies of many barred-spiral galaxies. Comparing our response models among themselves we find that the NGC~1300 morphology is best described by a thick disc model for the bar region and a 2D disc model for the spirals, with both components rotating with the same pattern speed $\Omega_p$=16~\ksk\!. In such a case, the whole structure is included inside the corotation of the system. The bar is supported mainly by regular orbits, while the spirals are supported by chaotic orbits.
\end{abstract}


\section{Introduction}
In \citet{kpg09} (hereafter PI) we have proposed three general models for the potential of the barred-spiral galaxy NGC~1300, which reflect three different geometries of the system.
In the extensive study presented in \citet{kpg09b} (hereafter PII) we have found several response models that reproduce at an acceptable degree morphological features of NGC~1300. Varying the pattern speed in models with all three assumed geometries, we concluded that the main features of NGC~1300 were better reproduced in models clustered around two $\Omega_p$ values, namely 16 and 22~\ksk\!. Taking into account the index used for the comparison of the models (PII) and the empirical by eye assessment of the resulting morphologies, we sorted out the following four ``best" cases. We remind that there was no single model reproducing all the features in an optimal way):
\begin{enumerate}
 \item \textbf{Model 1}: A 2D (Case A in PI) model, rotating with $\Omega_p$=22~\ksk\!. The main success of this model is the reproduction of the bar morphology.
 \item \textbf{Model 2}: A 2D (Case A in  PI) model, rotating with $\Omega_p$=16~\ksk\!. The main success of this model is the reproduction of the spiral morphology.
 \item \textbf{Model 3}: A thick disc model of cylindrical geometry (Case B in PI), rotating also with $\Omega_p$=16~\ksk\!. The main success of this model is a nice reproduction of the bar's morphology. Its response at the spiral region succeeds in giving the density maxima of the model roughly at about the same area occupied by the spiral arms of the galaxy. However, the response at the spiral region of the model forms a spurious ring structure.
 \item \textbf{Model 4}: Another thick disc model of cylindrical geometry (Case B in PI), rotating this time with $\Omega_p$=22~\ksk\! (as Model 1), develops a spiral structure that is worth to be presented as a separate model, since the spiral arms are formed through a different dynamical mechanism than that in Model 2.
\end{enumerate}

\begin{figure}
\begin{center}
\includegraphics[width=8cm]{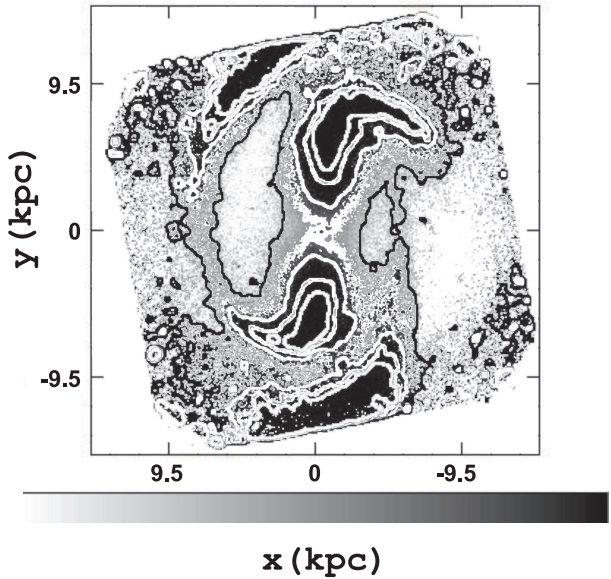}
\end{center}
\caption{The morphology of the barred-spiral perturbation in NGC~1300. The deprojected K-band image has been divided by the mean surface brightness at each radius. The original image is flipped in x and rotated clockwise by $4\pi/9$ to better compare with the orbital models. Isocontours of a smoothed K-band image are overplotted to emphasize the morphological details.}
\label{relima}
\end{figure}
We will follow the detailed orbital analysis of these four models in the subsequent sections. Nevertheless, we want to underline, that the goal of our paper is to present the dynamical mechanisms behind the observed structures. Besides the morphological similarities between response models and NGC~1300 we have chosen the four cases, because they represent different dynamical mechanisms leading to the formation of the bar and the spirals. This gives us the opportunity to compare different possible dynamics that lead to the support of similar structures in the galactic disc. In that sense the four models should be considered as typical cases for the description of different dynamics behind them.

A characteristic difference of the model potential from other \textit{barred-spiral} models we find in the relevant literature \citep{kc96, paq97, hk09, pkgb09} is its asymmetry. As we can see in the near-infrared image of NGC~1300 (fig.~3b in PI), the two spiral arms have different morphologies. Both arms emerge out of the ends of the bar as a continuation of an ansae structure. However, the ends of the ansae, i.e. the regions where the spiral arms begin, are not at equal distances from the center of the galaxy. The bar itself is asymmetric with its lower end being closer to the center of the galaxy.

Both spiral arms are strong at their beginning, close to the ends of the bar, in the near-infrared, even after correcting for star formation (PI). Notwithstanding, they weaken as we move azimuthally away from the bar. In order to clearly observe the exact shape of the spiral arms we divide in Fig.~\ref{relima} the deprojected K-band image with the mean surface brightness at each radius. The arm at the left side of Fig.~\ref{relima} could be considered as continuous, although its surface brightness varies azimuthally and has a clear minimum at a height corresponding roughly to the middle of the bar. However, the right arm in Fig.~\ref{relima} is conspicuously discontinuous and has a second strong part at the opposite end of the bar than the one it emerges from. All these asymmetries are taken into account in the potential we use (PI) and thus in our orbital calculations. We study a fully asymmetric dynamical behaviour.
We note that the spiral arms have in the K-band relative sharp features. This supports a quasi-stable character, since transient features, due to different wave packages, are more likely to be smooth.

In order to facilitate our calculations and the adaption of the potential models to our computer programs, we have flipped the potential around the central column and then we have rotated it by $4\pi/9$ counterclockwise. This allows us to study models in which the sense of rotation of the galaxy is counterclockwise and the major axis of the bar is close to the y-axis of our Cartesian coordinate system. All figures in the present paper have this orientation.
\begin{figure}
\begin{center}
\includegraphics[width=8cm]{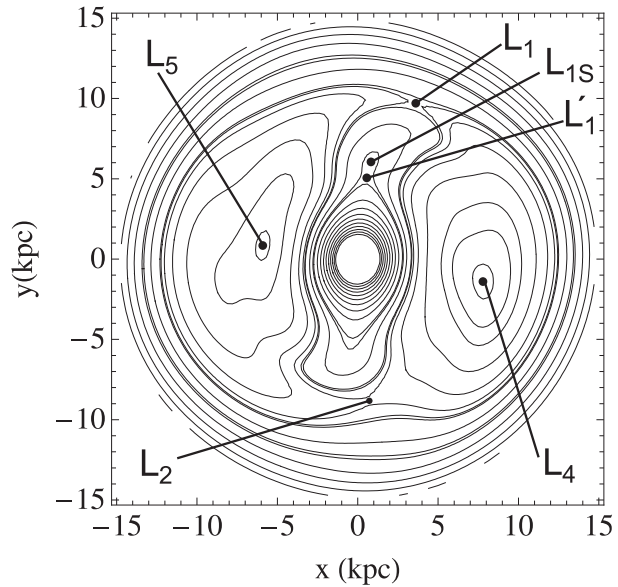}
\end{center}
\caption{The effective potential isocontours for our 2D Model 1 with $\Omega_p = 22$~\ksk\!. We have one stable (L$_{1S}$) and five unstable (L$_4$, L$_5$, L${_1}$, L$_{1'}$, L$_{2}$) Lagrangian points indicated with arrows.}
\label{effpot2d22}
\end{figure}
Equations of motion are derived from the Hamiltonian
\begin{equation}
H \equiv \frac{1}{2}\left(\dot{x}^{2} + \dot{y}^{2}\right) + \Phi(x,y) -
\frac{1}{2}\Omega_{p}^{2}(x^{2} + y^{2})=E_{J},
\end{equation}
where $(x,y)$ are the coordinates in the  Cartesian frame of reference as defined above,
rotating with angular velocity $\Omega_{p}$. $\Phi(x,y)$ is the
potential in Cartesian coordinates and can be anyone of the three models presented in PI. \ej is the numerical value
of the  Jacobi constant and dots denote time derivatives.

As we have seen in PII the range of $\Omega_{p}$ values for which a certain similarity between response models and galaxy could be traced was $12<\Omega_p<26$~\ksk\!. Beyond that range the resemblance of the response models with the morphology of the galaxy was visibly problematic. So we did not follow the evolution of the responses in models with pattern speeds out of this range.

\section{Model 1}
\begin{figure*}
\begin{center}
\includegraphics[width=16cm]{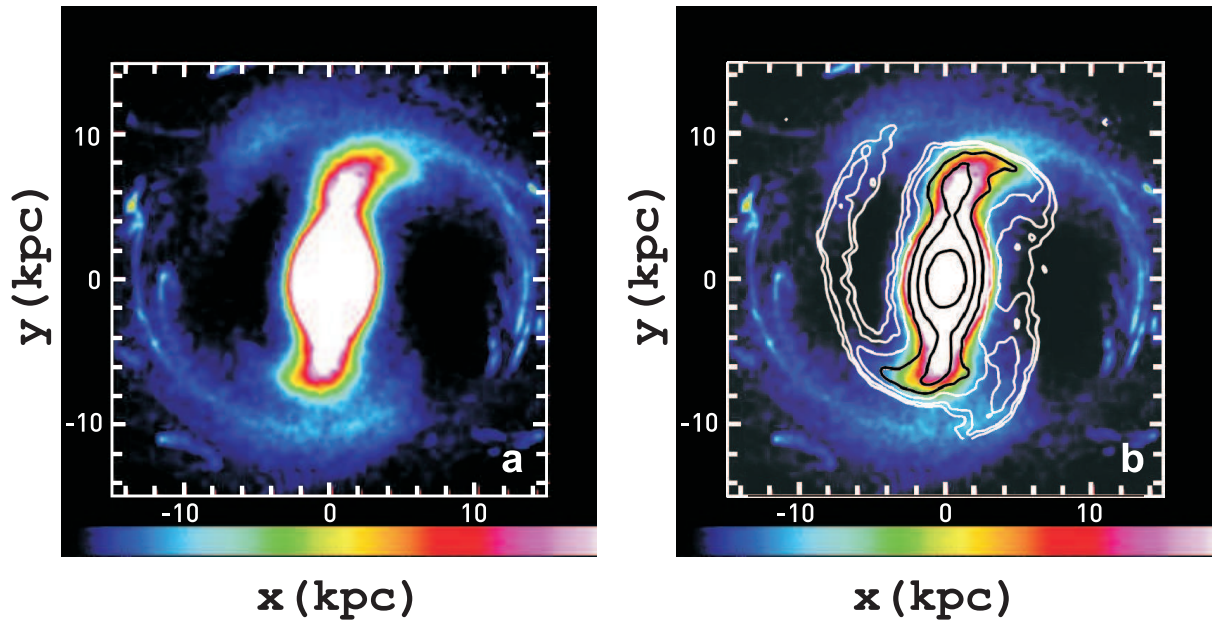}
\end{center}
\caption{Color-scaled density response of Model 1 (2D, $\Omega_p$ = 22~\ksk). (a) We observe the appearance of a typical grand-design barred-spiral morphology. (b) Characteristic isocontours of the deprojected K-band image of the galaxy, different than those given in Fig.~\ref{relima}, are overplotted on the response model. Model 1 has a bar with a morphology similar to the bar of NGC~1300, but its spirals are clearly more open than the spirals of the galaxy.}
\label{resp}
\end{figure*}
For $\Omega_p > 21$~\ksk the isocontours of the effective potentials have a remarkable shape. This is shown in Fig.~\ref{effpot2d22}, for our Model 1 ($\Omega_p = 22$~\ksk).
Besides the stable Lagrangian point at the center of the galaxy (not indicated on Fig.~\ref{effpot2d22}) and the usual equilibrium points at the sides of the bar (L$_4$, L$_5$), there appears one more stable Lagrangian point (L$_{1S}$), as a result of the complicated ``landscape" at the upper part of the bar. At this region we have {\em two} unstable Lagrangian points (L${_1}$ and L$_{1}^{\prime}$). L$_{1}^{\prime}$ appears close to the middle of the upper semi-major axis of the bar, at a lower \ej value ($-170759$) than that of L${_1}$ ($-167409$). Thus, between the unstable points L$_{1}^{\prime}$ and L${_1}$ appears L$_{1S}$, which is stable. Nevertheless, close to the lower end of the bar, we have only one unstable Lagrangian point, L$_{2}$, as usual (Fig.~\ref{effpot2d22}). Due to the asymmetry of the potential, L$_{2}$ has a larger \ej value ($-166797$) than that of L${_1}$\footnote{ Preliminary orbital calculations in potentials that have multiple Lagrangian points roughly along the major axis of the bar have been presented in the Padova meeting about ``Pattern speeds along the Hubble sequence'' in August 2008 \citep{pk09}. Also a study of the manifolds in a case with more than two Lagrangian point on the bar major axis has been presented by \citet{arg09}.}. We also note, that L$_4$ and L$_5$ in this case are unstable (see PI). However, as we will see below, the associated simple periodic orbits are stable in large intervals of the Jacobi constant.

To our knowledge it is the first time that a systematic orbital analysis is done in a galactic effective potential with more than two equilibrium points roughly along the major axis of the bar.

\begin{figure*}
\begin{center}
\includegraphics[width=12cm]{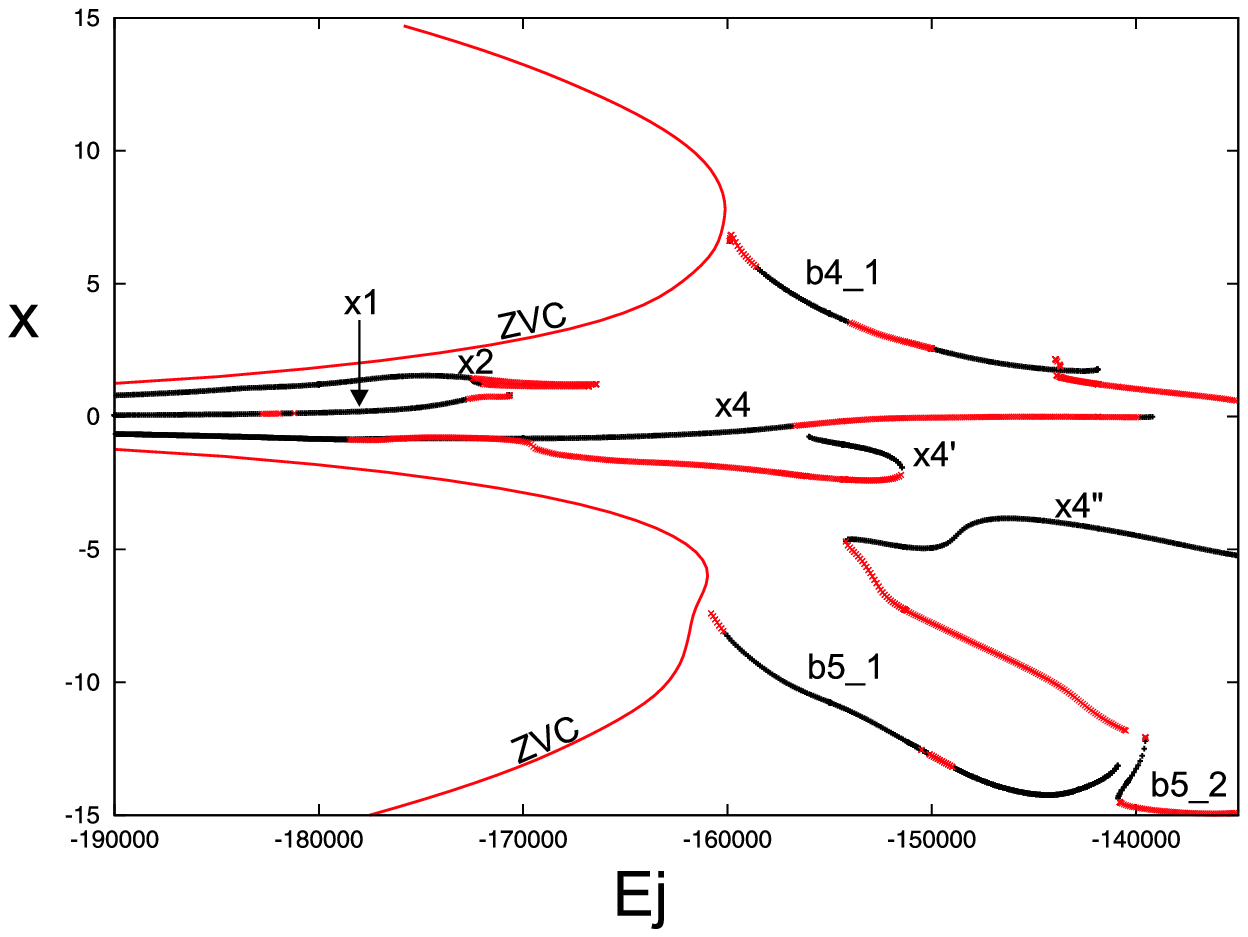}
\end{center}
\caption{($x$,\ej) characteristics of the main families of periodic orbits of Model 1. The curves designated with ``ZVC" correspond to the two branches of the zero velocity curve in this diagram. Black parts on the characteristic curves of the families indicate stability, while  grey (red in the online version) instability. The maxima of the ZVC's are near L$_4$, L$_5$. }
\label{char2d22}
\end{figure*}
\begin{figure*}
\begin{center}
\begin{tabular}{cc}
\hspace{-8mm}
\includegraphics[width=16cm]{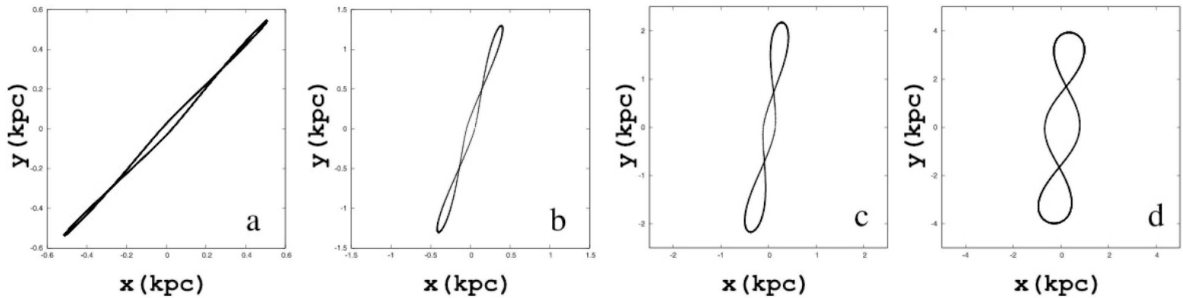}
\end{tabular}
\caption{Stable x1 orbits for Model 1 . (a) \ej = $-210000$, (b) \ej = $-190000$ (c) \ej = $-180098$ (d) \ej = $-170677.78$. The x1 orbit in (d) is the longest x1 along the major axis of the bar found in this model. The sizes of the frames are not equal in the four panels.}
\label{ox1}
\end{center}
\end{figure*}
The response of Model 1 is given in Fig.~\ref{resp}. It is a density map of the stellar response. In this and all subsequent density maps
of our response models the colored bar below the figure represents increasing density from left to right. Information about the initial set up and the initial conditions and other details about the construction of the models are given in PII. Here, with Fig.~\ref{resp}a we want to show that our 2D model rotating with $\Omega_p$ = 22~\ksk gets a barred-spiral morphology with spiral arms emerging close to the ends of the bar. The response morphology is typical of grand-design barred-spiral galaxies. The bar is a typical ansae-type bar, as the bar of NGC~1300, while the spiral arms are rather symmetric, and extend azimuthally roughly over an angle of $\pi/2$, beginning at the Lagrangian points L${_1}$ and L${_2}$ of the effective potential (cf. with Fig.~\ref{effpot2d22}).

In such a model it is tricky to point to a single $R_{CR}/R_b$ ratio (see figure 2 of PII). On one hand the ends of the bar are not at equal distances from the center of the galaxy. Then, $R_b$, the bar radius, is not unambiguously defined. On the other hand all Lagrangian points are found within a ring of width about 5~kpc and one cannot speak about a single corotation radius $R_{CR}$. By considering the upper length of the bar and the L$_1$ radius we find a $R_{CR}/R_b$=1.16, while we have $R_{CR}/R_b$=1.11 if we consider the lower length of the bar and the L$_2$ radius.

In Fig.~\ref{resp}b we overplot on the response Model 1, characteristic isophotes of the deprojected K-band image of the galaxy. The ansae-type response bar matches the size of the NGC~1300 bar both in length and width. The location of the ansae are in agreement with the location of the ansae of the bar of the galaxy. On the other hand the developed spiral structure of the model has no relation with the asymmetric spiral pattern we are trying to reproduce (Fig.~\ref{resp}b).
In order to understand the orbital content behind the observed structures, we firstly investigate the families of periodic orbits existing in Model 1. We want to check to what extent the observed structures are built by regular orbits trapped around stable periodic orbits used as building blocks. The standard way to build a bar in a rotating galactic potential is with regular orbits trapped around the stable x1 orbits \citep{gco80}.

\subsection{Families of periodic orbits}
The ($x$, \ej) characteristic for Model 1 is given in Fig.~\ref{char2d22}. The $\dot{x}$ initial condition of the orbits, is not taken into account in this diagram. Also we note that the L${_4}$, L$_{5}$ Lagrangian points are not exactly on the x-axis of the system. Thus, the upper and lower local maxima of the zero velocity curves (ZVC) at \ej $\lessapprox -160000$ do no correspond to L${_4}$ and L$_{5}$ respectively.
In Fig.~\ref{char2d22} we observe that the main part of the central family, x1, stops before \ej$=-170000$, while the other known family in rotating barred potentials, x2, together with a family bifurcated from it, reach a Jacobi constant value close to $-165000$. Our system has a Central Mass Concentration (CMC) term (PI), due to which both characteristics of x1 and x2 join at the center of the system (not included in Fig.~\ref{char2d22}). The two families evolve in parallel as the value of the Jacobi constant increases with x2 having always larger $x$ initial conditions than x1. Both families have large stable parts, which are drawn black, while the unstable parts along the characteristic curves are plotted in grey. For the estimation of the stability of the periodic orbits we calculate their H\'{e}non index \citep{h65}. The morphological evolution of the central family, which provides Model 1 with stable orbits extended along the major axis of the bar, resembles that of the propeller orbits presented by \citet{kp05}. They have loops already for small values of the Jacobi constant.

\begin{figure}
\begin{center}
\includegraphics[width=5cm]{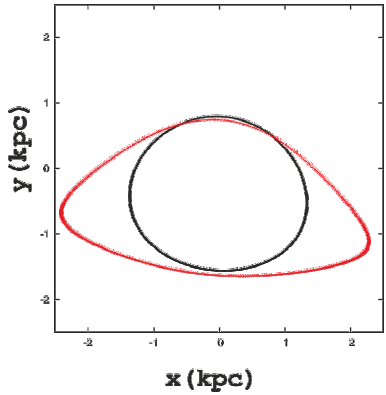}
\end{center}
\caption{Asymmetric, stable x2 periodic orbits in Model 1 for
\ej=$-180000$ (black) and for \ej=$-172572.1$ (grey - red in
online version).} \label{ox2}
\end{figure}

Fig.~\ref{ox1} shows this morphological evolution of x1. From (a) to (d) we plot stable representatives of this family
for \ej = $-210000$ (a), \ej = $-190000$ (b), \ej = $-180098$ (c) and  \ej = $-170677.78$ (d). The orbit given in Fig.~\ref{ox1}d is the longest stable x1.  The x2 orbits, extending in this model in the same Jacobi constant interval as the x1 family, are asymmetric. Two of them are given in Fig.~\ref{ox2} and both of them are stable. The black one is at \ej\!=$-180000$, while the grey one, for \ej=$-172572.1$ is the last stable x2 of the system. We observe that both of them are elongated along the minor axis of the bar as expected, they are asymmetric and their projections on the x-axis do not extend to radii larger than 2.5~kpc. Thus, they are embedded in the central region of Model 1 in Fig.~\ref{resp}. For Figs.~\ref{ox2} and all other figures with orbits in this paper the axes are considered as x, y Cartesian coordinates and the units on the axes are in kpc.

\begin{figure}
\begin{center}
\includegraphics[width=7cm]{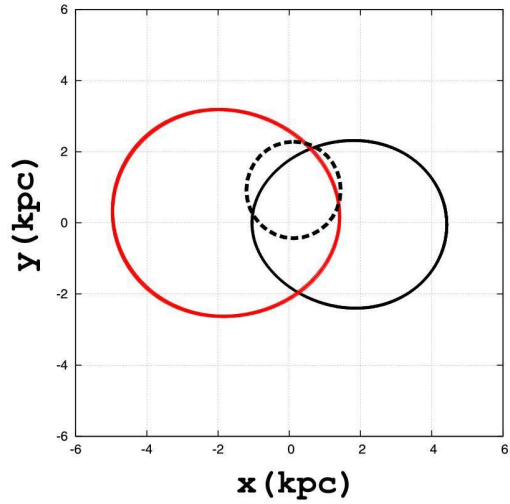}
\end{center}
\caption{Asymmetric with respect to the center of the system x4-like
orbits. x4 at \ej=$-170000$ (dashed); x4$^{\prime}$, a
bifurcation of x4, at \ej=$-154352.2$ (black); and x4$^{\prime
\prime}$ at \ej=$-150000$ (grey - red in the online
version)}. \label{ox4}
\end{figure}

\begin{figure}
\begin{center}
\includegraphics[width=7cm]{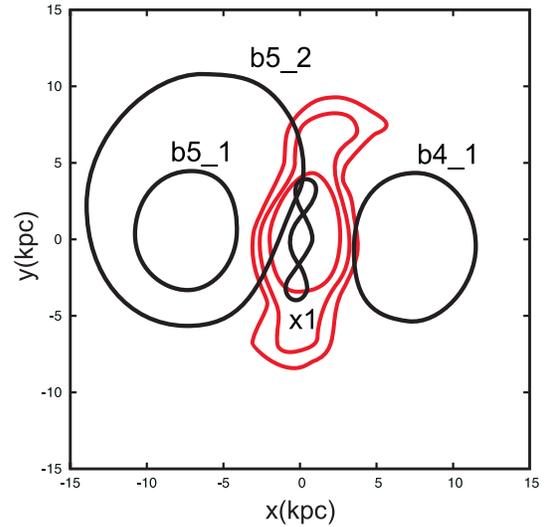}
\end{center}
\caption{Characteristic stable orbits of x1 (longest along the bar
in the system) and of the families b4\_1, b5\_1 and b5\_2, drawn
with black color, together with isodensity contours of Model 1
(grey curves - red in the online version). These orbits can
hardly support the developed morphology of Model 1.} \label{staborb}
\end{figure}

\begin{figure}
\begin{center}
\includegraphics[width=8cm]{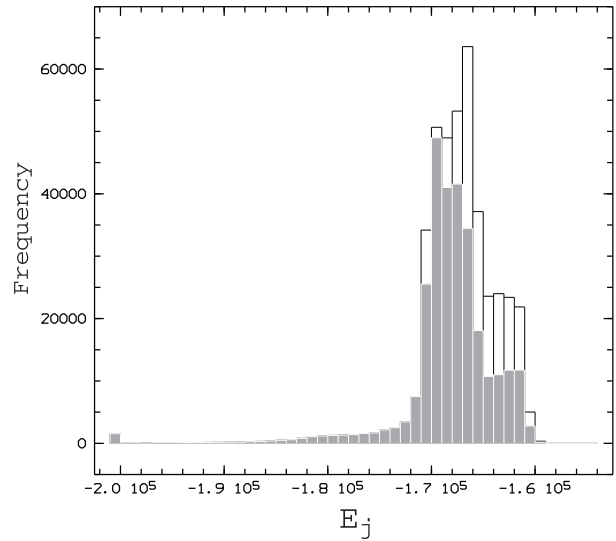}
\end{center}
\caption{Energy statistics for the particles of Model 1 depicted in
Fig.~\ref{resp}. Black refers to the total number of particles of
the model, while the overplotted grey histogram gives the
statistics at $r<9.2$~kpc, practically the statistics of the bar
particles.} \label{h2d22inout}
\end{figure}

Another set of periodic orbits that exist over large Jacobi
constant intervals in Fig.~\ref{char2d22} belong to x4 and x4-like
families. They are counterrotating orbits designated on
Fig.~\ref{char2d22} with x4, x4$^{\prime}$ and x4$^{\prime \prime}$.
All of them are displaced with respect to the center of the system.
Characteristic representatives of these families are given in
Fig.~\ref{ox4}. A grid is included in the figure for better
understanding the described asymmetries. The last main families of
stable periodic orbits in the system are families of banana-like
orbits around the Lagrangian points L${_4}$ and L$_{5}$. The
branches of the characteristic curves belonging to them are denoted
with b4\_1, b5\_1 and b5\_2 in Fig.~\ref{char2d22}. In order to have
an overview of the morphological features of Model 1 that could be
supported by stable periodic orbits, through trapping of regular
orbits around them, we plot in Fig.~\ref{staborb} orbits of the
families x1 (\ej = $-170677.78$), b4\_1 (\ej = $-154102$), b5\_1
(\ej = $-155000$) and b5\_2 (\ej = $-140398$), black curves,
together with isodensity contours of the bar of Model 1
(grey curves). The plotted x1 orbit is the one given in
Fig.~\ref{ox1}d, i.e. the longest stable orbit of this family. It is
clear that in this case stable x1 orbits can support only the
central bar region, as their projections on the major axis of the
bar cannot exceed radii of 4~kpc maximum. Thus, the loops of the x1
orbits are not related with the ansae of the response model. It is
evident from Fig.~\ref{staborb}, that the building blocks of the bar
of Model 1 are not the stable orbits of the x1 family. Moreover,
comparison of Fig.~\ref{staborb} with Fig.~\ref{resp} clearly shows
that the superposition of stable orbits of the banana-like families
can hardly support the open spiral pattern of the response model,
which in any case is not in good agreement with the spiral structure
of NGC~1300. In conclusion, the grand design barred-spiral structure
of Model 1 is rather unlikely to be shaped due to the usual
mechanism that uses stable periodic orbits as building blocks. The
question about what shapes the observed morphology remains.

\begin{figure*}
\begin{center}
\includegraphics{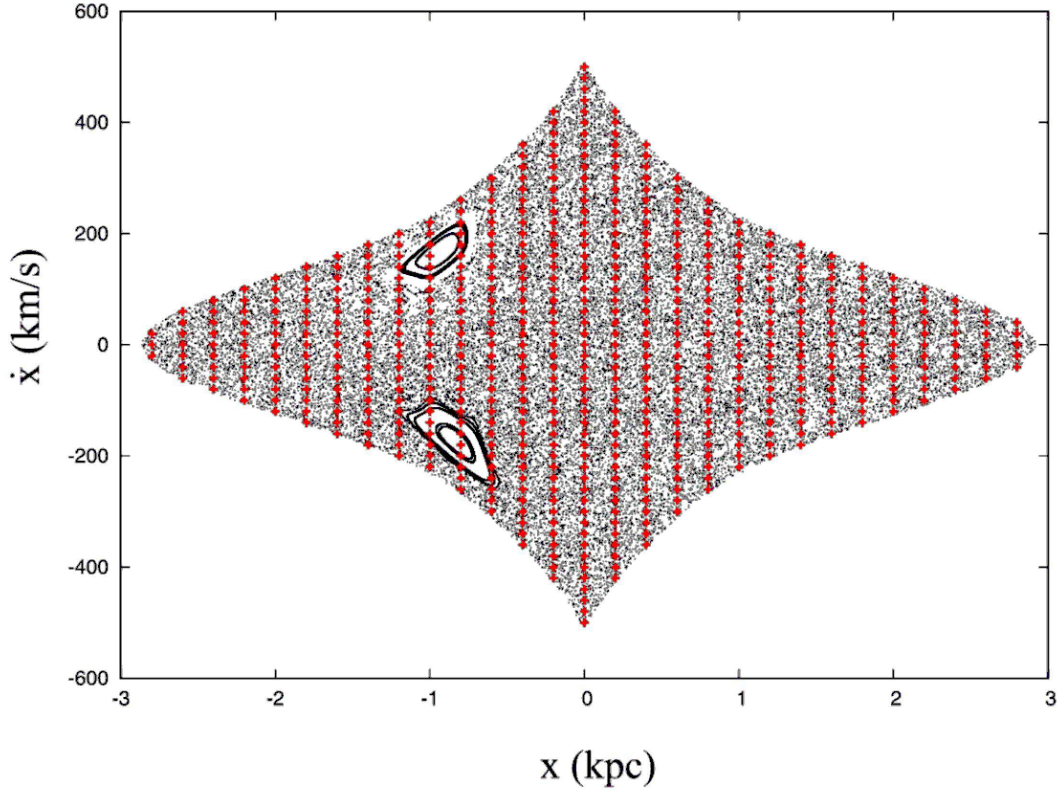}
\end{center}
\caption{The (x,\.{x}) surface of section at \ej = $-1.7\times 10^5
$ for Model 1. The overplotted red points indicate 587 initial
conditions that have been integrated for time corresponding to 7
pattern rotations. The morphology of the individual trajectories has
been examined in the text.} \label{sos170ic}
\end{figure*}

The next step that will help us understand the orbits of the
particles in the model is to study the statistics of their
Jacobi constants at several regions. In the present case it
is rather easy to roughly separate the particles on the bar from the
particles on the spiral arms, because the spirals of the model are
open and the region between spiral arms and bar is quite empty.
Essentially bar particles are those with $r<9.2$~kpc, while the vast
majority of the particles with $r>9.2$~kpc are on the spiral arms.

\begin{figure*}
\begin{center}
\begin{tabular}{cc}
\includegraphics[width=14cm]{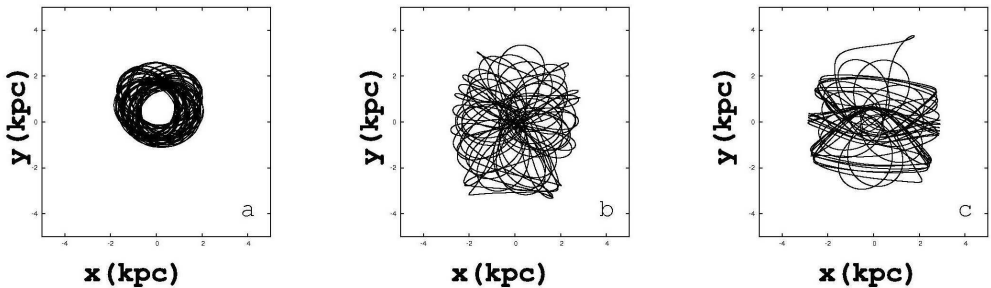}
\end{tabular}
\caption{Orbits for Model 1 at \ej = $-1.7\times 10^5 $. (a) A
sticky chaotic orbit close to x4, (b) a chaotic orbit staying inside
a 4~kpc radius and (c) a chaotic orbit that for a certain time
fraction follows a x2 flow. } \label{orbits170_no}
\end{center}
\end{figure*}
\begin{figure*}
\begin{center}
\begin{tabular}{cc}
\hspace{-8mm}
\includegraphics[width=170mm]{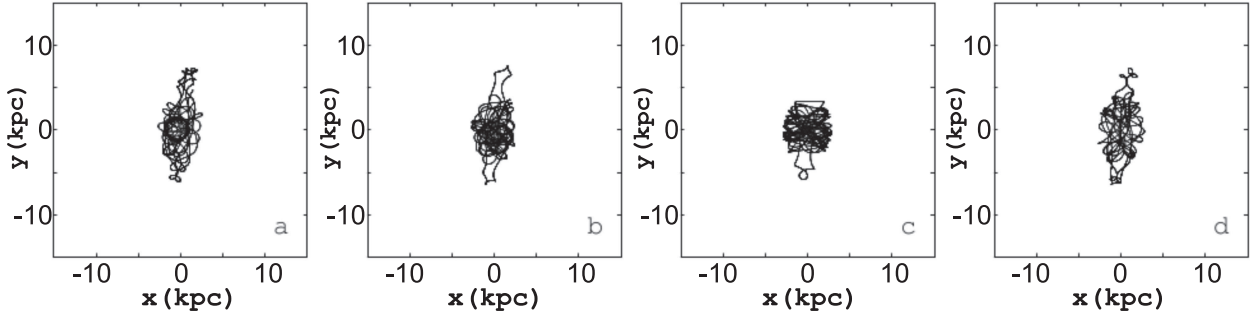}
\end{tabular}
\caption{Chaotic orbits in Model 1 with \ej=$-170000$, which
support the ansae bar morphology. From (a) to (d) the initial
$(x,\dot{x})$ conditions of the orbits are  $(-0.2,-360)$,
$(0.4,-320)$, $(1.3,-200)$ and $(0.8,-100)$.} \label{orb_ansae}
\end{center}
\end{figure*}

The histogram with the Jacobi constant statistics for Model
1 is given in Fig.~\ref{h2d22inout}. With grey is given the
statistics of the particles at $r<9.2$~kpc, while the black bars
that come on top of the grey ones refer to the total number
of particles and appear at higher levels beyond a certain value of
the Jacobi constant. We observe that we do not have any
particles for \ej $>-1.58\times 10^5$. By looking at
Fig.~\ref{char2d22} we realize that no particles are trapped around
stable banana-like periodic orbits. Only small segments of the
characteristics of b4\_1 and b5\_1 exist at \ej $<-1.58\times 10^5$.
Especially close to the upper local maximum of the ZVC, close to
L$_4$, there are only unstable b4\_1 periodic orbits. The interval
\ej $<-1.58\times 10^5$ includes only a small segment of stable
periodic orbits belonging to the b5\_1 banana-like family.
Practically the characteristics at the right part of
Fig.~\ref{char2d22} belong to families that are not related with the
model. The grey histogram, that refers to the bar
particles, shows that these particles have Jacobi constants
in the interval $-1.71\times 10^5 <$ \ej $<-1.65\times 10^5$, with a
peak at $-1.7\times 10^5 <$ \ej $<-1.69\times 10^5$. What is the
orbital behaviour at these Jacobi constants? The range of
the Jacobi constants roughly covers the region between the
last x1 orbits and the local maxima of the ZVC close to the L$_4$
and  L$_5$ points (Fig.~\ref{char2d22}). Knowing the range of \ej's
that contributes mostly to the orbital content of the bar, we
investigate the surfaces of section at these Jacobi
constants. The (x,\.{x}) surface of section for \ej = $-1.7\times
10^5 $, close to the mode of the distribution, is given in
Fig.~\ref{sos170ic}. There are depicted about $5\times 10^4$
consequents with black points. The only two conspicuous islands of
stability belong to the asymmetric x4 retrograde orbits. The one
with negative velocity is like the dashed orbit in Fig.~\ref{ox4},
while the other one, with positive velocity, is its symmetric with
respect to the y=0 axis. Apart from these two islands of stability
the rest of the surfaces of section is dominated by Chaos. We know
however, that at this \ej we encounter most particles that
contribute to the bar structure.

\begin{figure}
\begin{center}
\includegraphics[width=8cm]{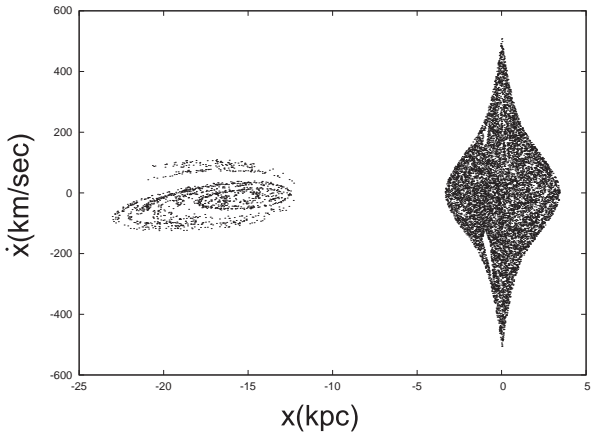}
\end{center}
\caption{The (x,\.{x}) surface of section for \ej=$-1.67\times 10^5$
in Model 1.} \label{sos167}
\end{figure}

The chaotic sea can be in principle generated by a single chaotic
orbit, if we integrate it for a long time. However, starting with
different initial conditions and by integrating for relatively short
times we do not obtain necessarily similar orbital morphologies.
Integrating for finite time intervals a large number of initial
conditions we simulate the situation of following the trajectories
of individual particles on the galactic disc, and this allows us to
understand whether they support a specific morphology within this
time interval. The red symbols overplotted on the surface of section
indicate 587 initial conditions on the (x,\.{x}) surface of section
in Fig.~\ref{sos170ic}, that have been integrated for time
corresponding to 7 pattern rotations. We will refer to all of the
resulting trajectories also as ``orbits", despite the fact that the
chaotic orbit is essentially one. The supported morphologies we
found were quite distinct, and this allowed us to proceed to a crude
by eye classification as follows:

Around 7\% of the trajectories were related with the x4 retrograde
family, being either regular quasi-periodic orbits trapped around
them, or related sticky chaotic orbits. A typical orbit of this kind
is given in Fig.~\ref{orbits170_no}a. Another 40\% of the integrated
trajectories were chaotic, staying in the inner bar region without
ever exceeding a radius r=4~kpc. They were essentially supporting
the part of the bar without the ansae. Such an orbit is presented in
Fig.~\ref{orbits170_no}b. Only 1\% of the chaotic orbits could be
associated with x2-like orbits (Fig.~\ref{orbits170_no}c, cf. with
Fig.~\ref{ox2}). The remaining 52\% of the integrated initial
conditions belong to orbits in the chaotic sea, which share a common
feature. These particles, after spending part of the integration
time in the central bar region as in Fig.~\ref{orbits170_no}b, they
visit the immediate neighborhood above and below the central part
contributing to the formation and support of the ansae. Such orbits
are presented in Fig.~\ref{orb_ansae}. We note that L$_{1}^{\prime}$
appears at \ej=$-170759$, very close to the \ej value of the surface
of section we study, indicating that the appearance of this type of
orbits is associated with the presence of the stable and unstable
manifolds of the unstable L$_{1}^{\prime}$ equilibrium point. This
type of \textit{chaotic} orbits give the strong ansae type
morphology of the bar of the response Model 1.

The other morphological feature, that characterizes Model 1, are its
open spirals (Fig.~\ref{resp}). Despite the fact, that they do not
match the NGC~1300 spirals, they are a strong feature of its
grand-design morphology. They consist of particles at $r>9.2$~kpc.
As we can conclude from Fig.~\ref{h2d22inout} the mode, the value
that occurs most frequently in the distribution of these particles,
is found for $-1.68\times 10^5 <$ \ej $<-1.66\times 10^5$. In order
to find the orbital content of the model spirals we follow the same
procedure. The (x,\.{x}) surface of section for \ej=$-1.67\times
10^5$  is presented in Fig.~\ref{sos167}. Being closer to the local
maxima of the ZVC (Fig.~\ref{char2d22}) there appears in the diagram
also a region to the left of the central area of the surface of
section. At the central region the only conspicuous islands of
stability are again those belonging to x4, as in the previous case
we studied at \ej=$-1.7\times 10^5$.

\begin{figure*}
\begin{center}
\includegraphics[width=\textwidth]{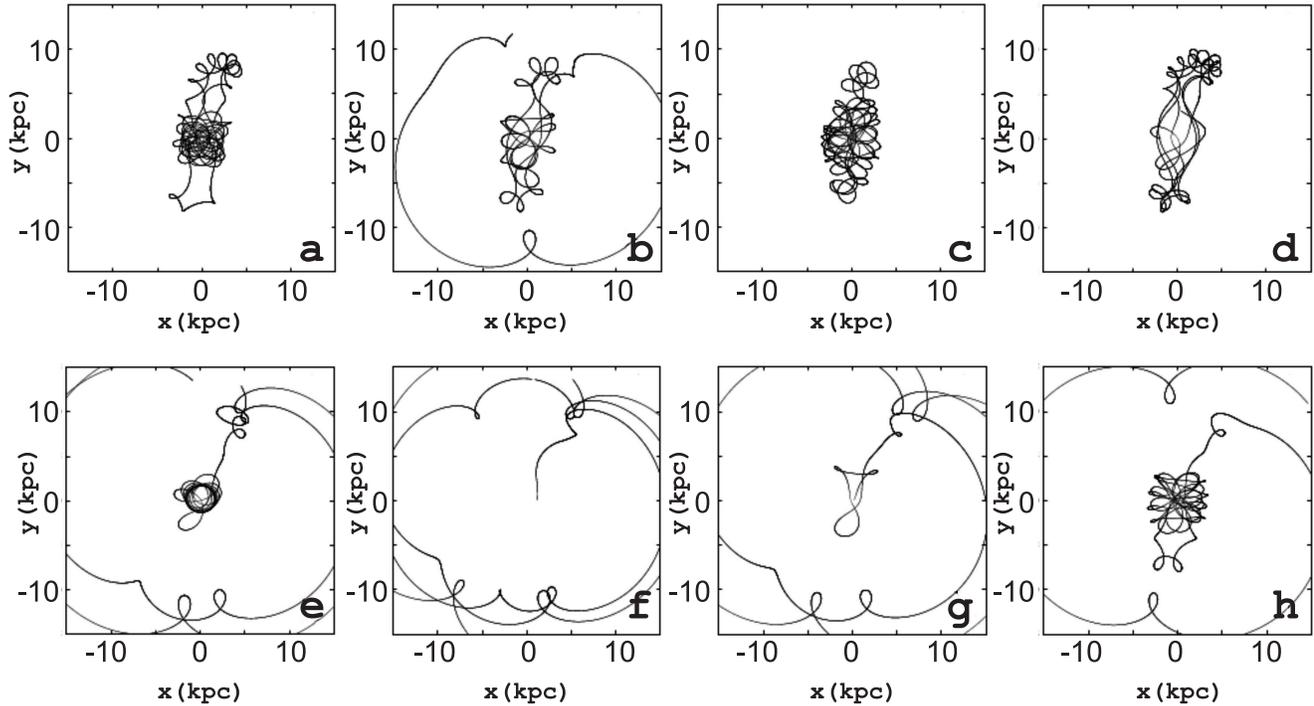}
\end{center}
\caption{Chaotic orbits at \ej=$-1.67\times 10^5$ in Model
1. 63\% of the initial conditions we integrated support a bar
morphology with ansae, while 26\% contribute to the formation of the
spiral structure of the model. From (a) to (h) the initial
$(x,\dot{x})$ conditions of the orbits are: $(-0.4,-360)$,
$(0.2,-360)$, ($ 0.6,-280)$, ($-1.8,-160$), ($-1.4,-160$), ($1.2,
40)$, ($ 0.2, 200$), ($0.6, 240$).} \label{orb167-2d22}
\end{figure*}

We have integrated again a large number of initial conditions for a
time equal to 7 pattern rotations. In this case the number of orbits
that support a bar with ansae increased to 63\%, while an additional
26\% abandon the bar region and by going through the L$_1$ location
visit the area of the disc and the spirals. Typical orbits of the
kind we describe can be observed in Fig.~\ref{orb167-2d22}. The
mechanism that shapes the open spirals of Model 1 is based on the
presence of the family generated at the unstable Lagrangian point
L$_1$, which has \ej=$-167409$. For a Jacobi constant just
greater than it, at \ej=$-167000$, which is the Jacobi
constant of the surface of section we study, the unstable periodic
orbit associated with the open spiral pattern already exist. The
chaotic sea is structured by the presence of the manifolds of this
unstable periodic orbits. For this particular response model, the
mechanism that builds its spiral arms is the one valid also for the
spirals of NGC~4314 \citep{p06}. This mechanism, as in the case of
NGC~4314, creates spirals with a strong part, that reaches
azimuthally angles up to $\pi/2$ starting from the ends of the bar.
Details about how chaotic spirals can be formed this way can be
found in \citet{vsk06}, \citet{rmag06}, \citet{vte07},
\citet{tkec09}, \citet{gco09} (see also articles by the same authors
in \citet{cia08} and references therein). Our analysis shows
however, that the spirals of NGC~1300 cannot be related with this
mechanism. Thus, we continue the investigation of our response
models in order to find a better explanation of the dynamical origin
of the NGC~1300 spirals. We remind though, that the Model 1 bar
reproduces in a satisfactory way the bar of NGC~1300.

\section{Model 2}
\begin{figure*}
\begin{center}
\includegraphics[width=16cm]{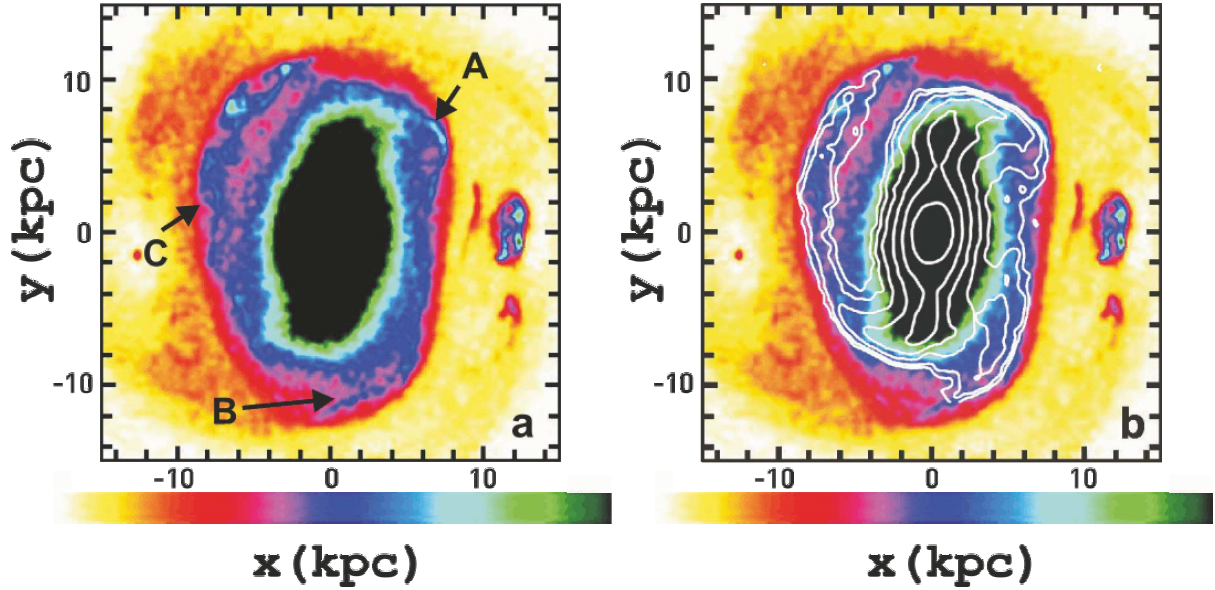}
\end{center}
\caption{The colored-scaled density response of Model 2 (2D with
$\Omega_p = 16$~\ksk). In (a) we point with arrows to the three
morphological features that resemble the corresponding structures of
NGC~1300. In (b) we overplot on the model isophotes of the
deprojected K-band image of NGC~1300, that show the level of
agreement of the spiral morphology between model and galaxy.}
\label{resp2d16}
\end{figure*}

In PII we have seen that by lowering the pattern speed of the system
in the 2D case to $\Omega_p = 16$~\ksk we obtained an excellent
matching of the location of the spirals between model and galaxy.
This case is our Model 2 of the present paper. We summarize its
morphology with the help of Fig.~\ref{resp2d16}. There are three
main successes of Model 2 in its comparison with the NGC~1300
spirals: (i) the abrupt break of the upper arm, just after it
emerges from the upper end of the bar. This is indicated in
Fig.~\ref{resp2d16}a with an arrow labeled ``A". (ii) the spiral
fragment at the lower right side of the bar. An arrow labeled ``B"
points in Fig.~\ref{resp2d16}a to this feature. (iii) the location
and the pitch angle of the rather continuous left arm, indicated
with the arrow labeled ``C".  The overplotted on the model isophotes
of the deprojected K-band image, show clearly the agreement of these
features (Fig.~\ref{resp2d16}b). This time the spiral structure is
modeled at a very satisfactory level. However, the only feature of
the bar that agrees with the NGC~1300 bar is its size. The model bar
is thicker than the one of the galaxy and has only a weak ansae
character.

\begin{figure}
\begin{center}
\includegraphics[width=8cm]{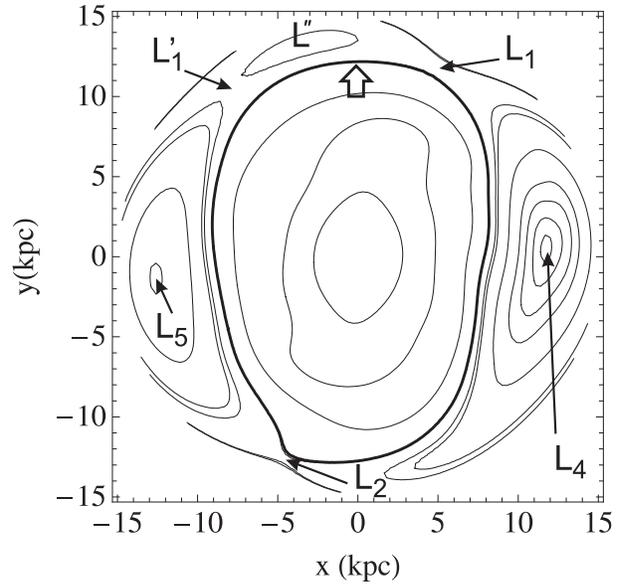}
\end{center}
\caption{The effective potential isocontours for our Model 2 with
$\Omega_p = 16$~\ksk\!. We have two stable (L$_4$, L$_5$) and four
unstable (L${_1}$, L$_{1}^{\prime}$, L$_{1}^{\prime\prime}$,
L$_{2}$) Lagrangian points indicated with black arrows. The white
thick arrow points to an isocontour, that surrounds the bar-spiral
morphology.} \label{effpot2d16}
\end{figure}

Particularly interesting is the comparison of the effective
potential isocontours (Fig.~\ref{effpot2d16}) with the resulting
response model. Now, for $\Omega_p = 16$~\ksk, the Lagrangian points
have moved to larger distances from the center. We have the usual
two stable L${_4}$ and L${_5}$  equilibrium points (at approximately
$-149690$ and $-150300$) and the two unstable L${_1}$ and L$_{2}$ at
$-152960$ and $-153020$ respectively. Besides them there is a third
unstable equilibrium point, L$^{\prime}{_1}$, at $\approx (-8,10)$.
As a consequence, there is one more equilibrium point,
L$^{\prime\prime}{_1}$, between L${_1}$ and L$^{\prime}{_1}$, which
in this case is unstable. All Lagrangian points appear beyond the
end of the barred-spiral morphology of the response model
altogether. Contrarily to Model 1, all the Lagrangian points
in Model 2 are within a relative narrow ring of width about 1.8~kpc.
However, the $R_{CR}/R_b$ ratios we find deviate more among
themselves. By considering the upper length of the bar and the L$_1$
radius we find this time $R_{CR}/R_b$=1.44, while we have
$R_{CR}/R_b$=1.68 if we consider the lower side of the bar and the
L$_2$ radius. 

We underline the presence of an isocontour in the effective
potential, which we have drawn with a thicker line and to which we
point with a thick white arrow. There is a striking similarity
between the shape and the location of this isocontour in
Fig.~\ref{effpot2d16} and the border of the red region that
surrounds the barred-spiral structure of the response Model 2 in
Fig.~\ref{resp2d16}. We discuss the role of this isophote for the
dynamics of Model 2 in the following paragraphs.

\begin{figure*}
\begin{center}
\includegraphics[width=14cm]{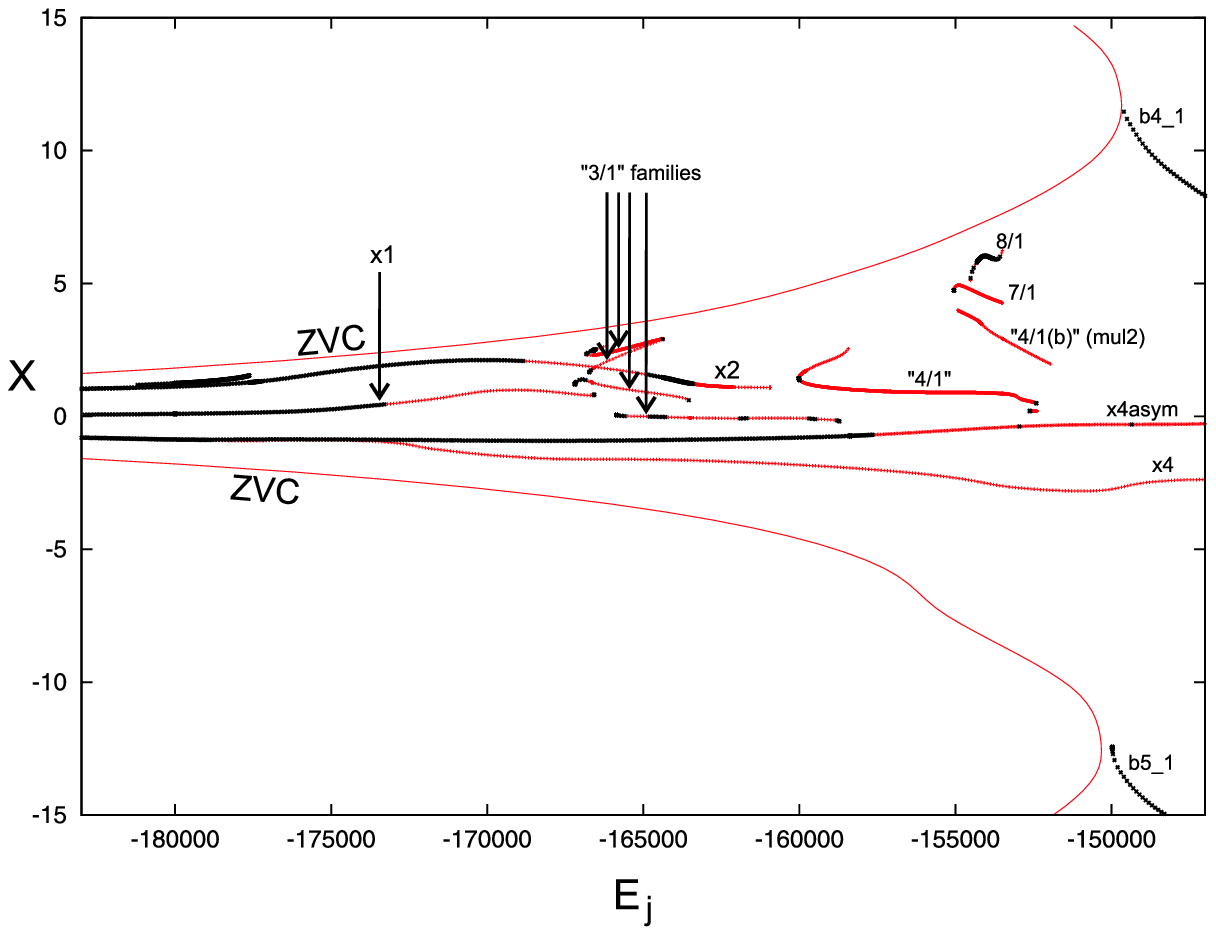}
\end{center}
\caption{The characteristics of the main families in Model 2 (2D
with $\Omega_p = 16$~\ksk). We note the presence of the 4/1, 7/1 and
8/1 families at Jacobi constants $-160000 \succapprox$ \ej
$\succapprox -150000$. Black segments of the characteristics
indicate stability, while grey (red in the online version)}
instability. The maxima of the ZVC's are near L$_4$, L$_5$.
\label{char2d16}
\end{figure*}

The characteristics of the main families of periodic orbits in an
(\ej, $x$) diagram, similar to the one presented in
Fig.~\ref{char2d22} for Model 1, is given in Fig.~\ref{char2d16}.
The evolution of the x1, x2 families is similar to that of Model 1,
but the Jacobi constant interval between the last stable x1
and the corotation region has increased significantly. Between them
we have calculated segments of characteristics belonging to families
with asymmetric morphologies we have classified as 4/1, 7/1 and 8/1.
We also note the presence of segments of several 3/1 type families
close to the largest Jacobi constants, where we still encounter x1
and x2 orbits. At all Jacobi constants in
Fig.~\ref{char2d16} we have always orbits belonging to x4 and its
bifurcations. It is beyond the scope of the present paper to study
the interconnections between the various families.

\begin{figure*}
\begin{center}
\includegraphics[width=18cm]{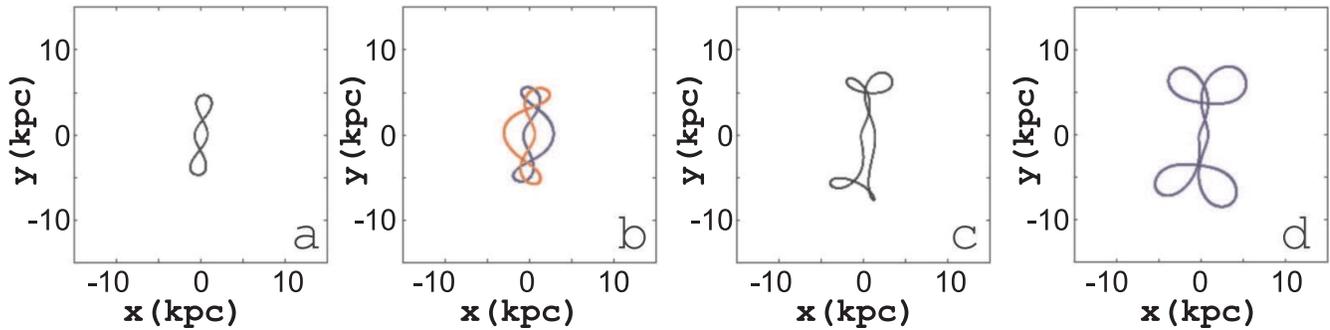}
\end{center}
\caption{Periodic orbits in Model 2, that are elongated along the
major axis of the bar. The orbits in (a), (b) and (c) are stable,
while the orbit in (d) is unstable.} 
\label{pobar2d16}
\end{figure*}

This time we have found periodic orbits extended roughly along the
major axis of the system, which could contribute to the bar
structure. In Fig.~\ref{pobar2d16} we present some of them
independently of their stability. (a) The longest stable x1 orbit at
\ej = $-166579.83 $. It has loops that now reach distances about
5~kpc from the center, (b) two 3/1 orbits at $-164476.07$ (dark
grey) and $-163642.8$ (light grey), that are close to symmetric with
respect to the y-axis and together can be considered as bar
supporting, (c) a 4/1 orbit from the tiny stable part of the
characteristic of this family at \ej=$-160018.42 $ and finally at
(d) the 4/1 representative at \ej=$-153480$.

\begin{figure}
\begin{center}
\includegraphics[width=8cm]{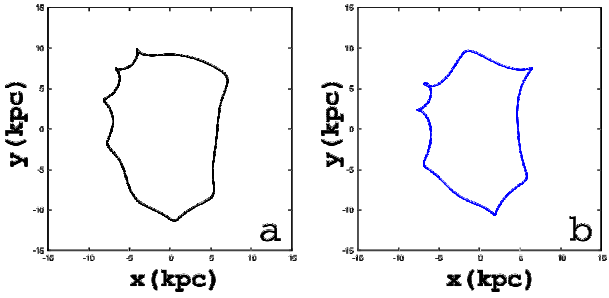}
\end{center}
\caption{Orbits of the families 8/1 (a) and 7/1 (b) in Model 2.}
\label{pospir2d16}
\end{figure}

\begin{figure}
\begin{center}
\includegraphics[width=8cm]{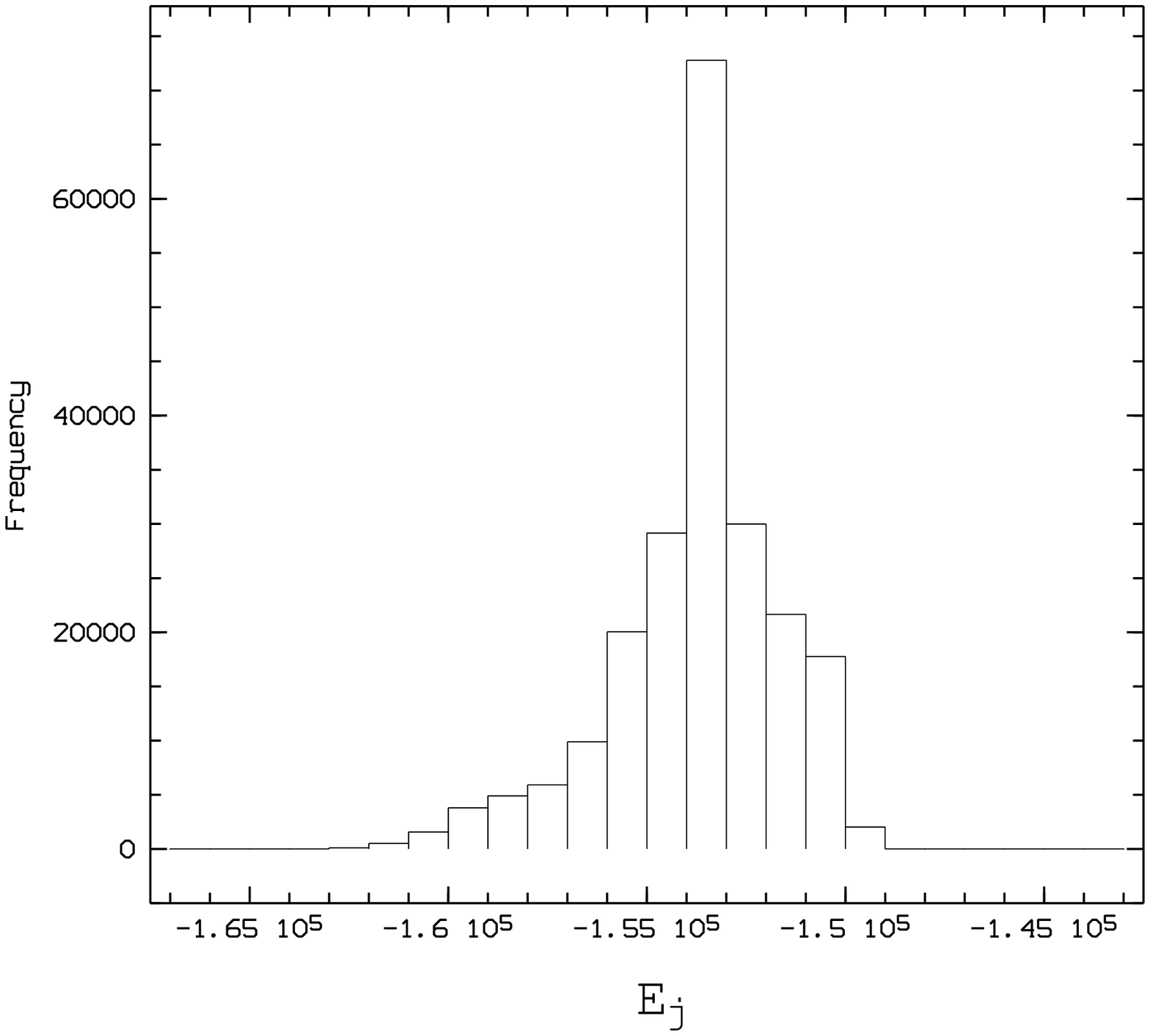}
\end{center}
\caption{Histogram that presents the Jacobi
constant statistics of the particles in an isolated region that
includes the spirals, but not the bar, in the response Model 2.}
\label{histoedw}
\end{figure}

We also give the morphology of the two families above the 4/1, at
larger $x$. In (a) a stable 8/1 (\ej=$-154280.23$) and in (b) an
unstable 7/1 (\ej=$-155046.68$) (Fig.~\ref{pospir2d16}). In
order to trace the orbits of the particles that successfully
reproduce the spiral structure of NGC~1300 in Model 2, we follow the
same procedure as in Model 1. We isolate an area around the bar of
the response model, which includes the spiral arms and we calculate
the Jacobi constants of the particles in this area. The
histogram with the Jacobi constant statistics at this
region is given in Fig.~\ref{histoedw}. The histogram shows that
the peak happens at the bin $-154000<$ \ej $ < -153000$, while
practically the Jacobi constants of all particles are
$-156000<$ \ej $ < -150000$.  Most particles on the spirals of Model
2 have these \ej values. By looking at Fig. ~\ref{char2d16} we
realize that at this region exist the 7/1 and 8/1 families, the
``4/1" family with double multiplicity, as well as a segment of the
4/1 characteristic. Only for the 8/1 family the H\'{e}non index is
$|\alpha| < 1$ (black segments in Fig.~\ref{char2d16}. This analysis
shows that whatever the dynamical mechanism behind the appearance of
the spiral structure is, it is related with particles on
trajectories inside corotation.

\begin{figure*}
\begin{center}
\includegraphics[width=16cm]{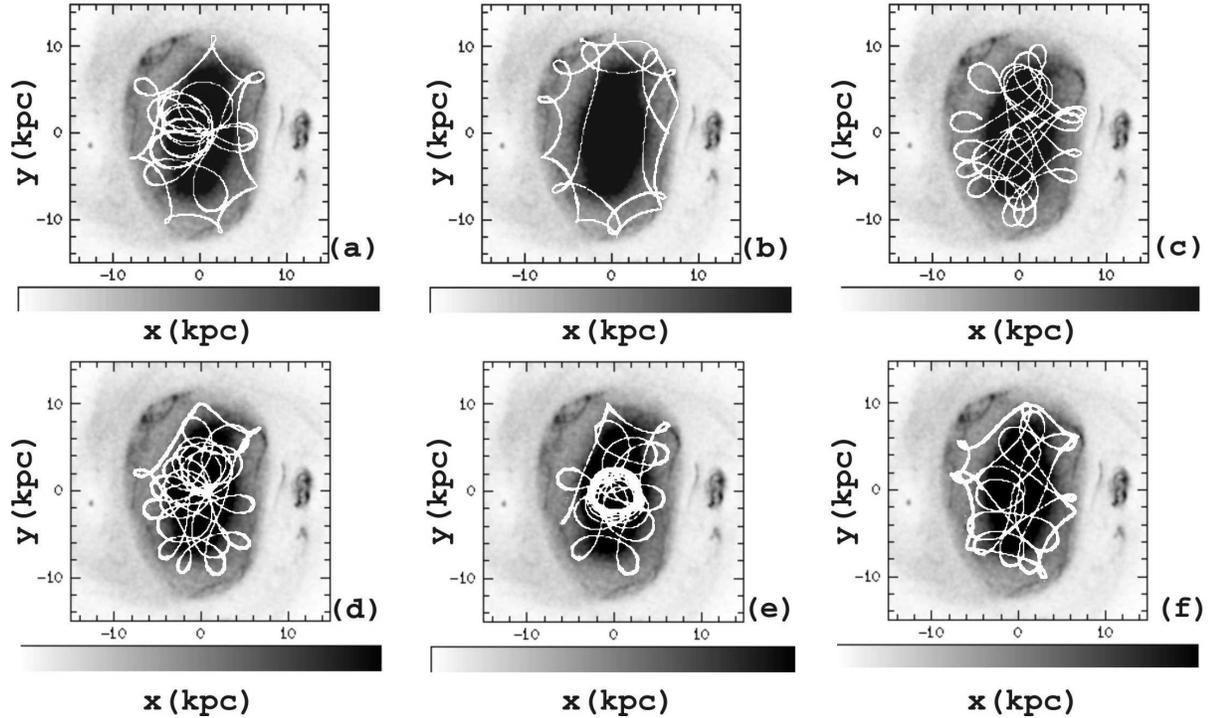}
\end{center}
\caption{Typical spiral supporting orbits in Model 2 with
\ej=$-153500$ in the upper row (panels (a), (b) and (c)) and with
\ej=$-155000$ in the lower row (panels (d), (e) and (f)). The
spirals are supported by the loops of the orbits at their region. At
the background we give the response Model 2.} \label{np2d16_1535}
\end{figure*}

On the (x,\.{x}) surface of section for \ej=$-153500$ there
are no islands of stability discernible. For this Jacobi
constant all families of periodic orbits we mentioned existing at
the $-156000<$ \ej $ < -150000$ interval are unstable. Again in this
case we integrated a large number of initial conditions on a grid
for time equal to seven pattern rotations.
The inspection of these orbits helped us realize that the vast majority of the chaotic orbits during the integration time have loops at their apocentra almost on the thick isocontour of the effective potential drawn in Fig.~\ref{effpot2d16}. The red region surrounding the barred-spiral structure in Fig.~\ref{resp2d16} is formed by the same orbits. As these orbits perform their loops, they stay longer time at the region of the heavy isopotential in Fig.~\ref{effpot2d16} and this is the way the local density maximum corresponding to the spiral arms are formed. In Fig.~\ref{np2d16_1535} we give six typical spiral supporting orbits in Model 2 with \ej=$-153500$ (upper row) and with \ej=$-155000$ (lower row). We plot them having as background the response of the model in order to clearly see the relation between orbits and response morphology. The initial $(x,\dot{x})$ conditions are $(0.6,-297.5)$ (a), $(6.6,-50)$ (b), $(-2,2 170)$ (c) and $(0.05,-407)$ (d), $(0.05,-270)$ (e) and $(-4.15,-77.5)$ (f).
We have practically the same dynamical behaviour at every Jacobi constant in the interval $-156000<$ \ej $ < -150000$. Orbits of similar morphology are found by integrating initial conditions at all these surfaces of section, and contribute to the formation of the spiral pattern. The only difference is in the mean radius of these chaotic orbits, which depends on the Jacobi constant. Their existence over a $\Delta$\ej range give the observed thickness of the model spirals.

The reason for which we have the formation of an open spiral and not
a ring is that the orbits that participate in the formation of the
structure are chaotic. Longer stability intervals along the
characteristics of the involved families would trap a large number
of quasi-periodic orbits around them leading to the formation of
rings on the response models. Even the size of the islands of the
stable 8/1 orbits are tiny, so the amount of trapped orbits is
small. However, despite the fact that the stable periodic orbit
influences a small part of the phase space and the fact that apart
from the 8/1 family all other periodic orbits are unstable,
we observe a similarity between the morphology of the
existing periodic orbits (see Fig.~\ref{pospir2d16}) and the chaotic
orbits that support the spiral structure (Fig.~\ref{np2d16_1535}).
This dynamical behaviour is similar to the one found by
\citet{paq97} for the orbits that support the outer envelope of the
bar in the potential of NGC~4314.

Model 2 gave an acceptable solution for the spirals of NGC~1300.
However, out of the morphological features of the bar only its size
matches the size of the NGC~1300 bar. The combination of the two
acceptable solutions, namely Model 1 for the bar and Model 2 for the
spirals, in a single model with two pattern speeds is problematic. A
careful inspection of the NGC~1300 morphology shows that a bar
rotating with a different pattern speed than the spirals will after
a certain time visit the spiral region and the combined structure
will change drastically in time. A way out of this dilemma is
offered by our next model we have chosen to present, i.e Model 3.

\section{Model 3}
The two models we have presented up to now have a planar geometry.
For models in the general class of thick disc models (PI), we have
realized in PII, that in the range of $\Omega_{p}$ around
16~\ksk$\!,$ the bar was becoming considerably thinner than the
corresponding 2D models with the same pattern speed. In this section
we describe the orbital behaviour in a thick disc model (see PI),
rotating with $\Omega_{p}$=16~\ksk\!. Around this $\Omega_{p}$ value
the $d_{GH}$ index in PII improves in models, where we weight the
Jacobi constants we use in our models. However, the
thinning of the bar and the enhancement of the density at the ends
of the bar that emphasizes its ansae character, is a basic feature
of all thick disc models with relatively low patterm speeds. The
best similarity with the NGC~1300 image is observed at
$\Omega_{p}$=16~\ksk$\!$ (see Fig.~3 in PII). The
$R_{CR}/R_b$ ratios we find considering the L$_1$ and L$_2$
Lagrangian points are this time 1.5 and 1.7 respectively.

\begin{figure}
\begin{center}
\includegraphics[width=8cm]{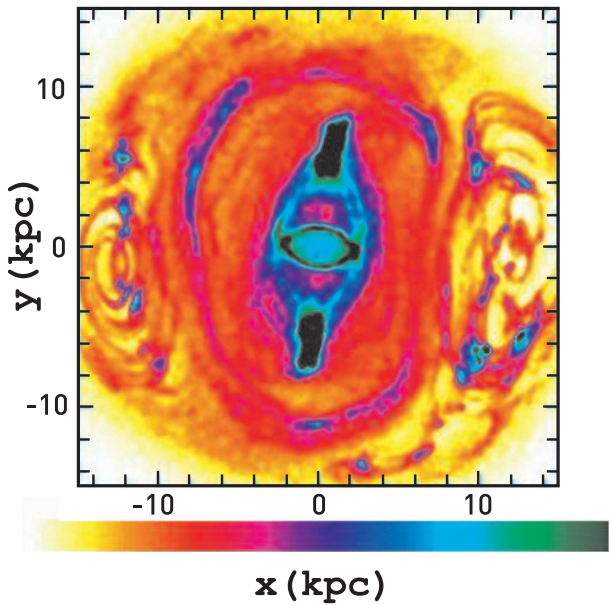}
\end{center}
\caption{The color-scaled density response of our Model 3, a thick
disc model rotating with $\Omega_{p}$=16~\ksk\!.} \label{respo3Dd16}
\end{figure}

The response of our Model 3 is given in Fig.~\ref{respo3Dd16}. The feature that appears considerably improved compared
with Model 2 is the bar. The length of the bar remains the same, however the ends of the bar appear thinner and elongated at the region which corresponds to the ansae of the NGC~1300 bar (see also the corresponding section in PII). On the other hand less successful is the model in describing the spiral of the galaxy. At the same region, where we obtained the spiral arms in the 2D Model 2, now a ring is formed. The density along this ring is not constant, but the resulting feature does not match the spiral of the galaxy as good as the model with the planar geometry (cf. Figs.~\ref{resp2d16} and \ref{respo3Dd16}). Below we describe the differences in the orbital dynamics between the two models, that are reflected in the morphology of their responses.

\begin{figure*}
\begin{center}
\includegraphics[width=10cm]{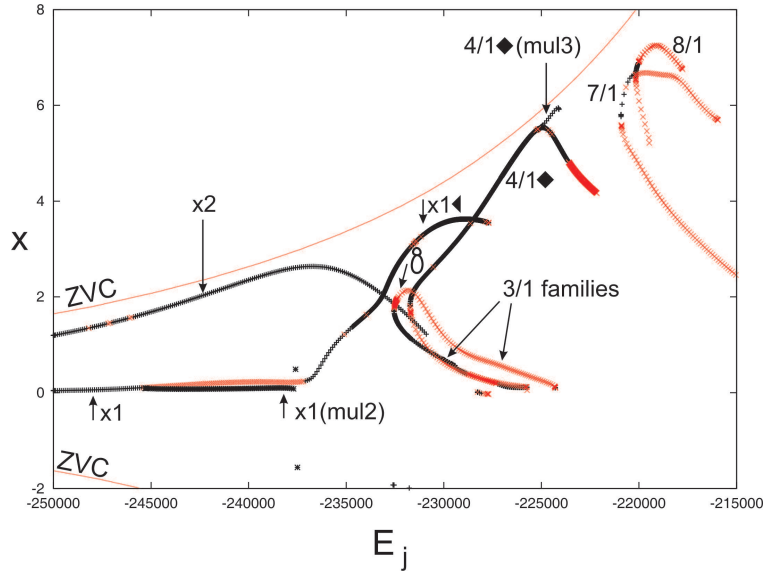}
\end{center}
\caption{The (\ej, $x$) characteristics of the main families of
periodic orbits of Model 3. Black indicates stable orbits, while
grey (red in the online version)} unstable.
\label{char3Dd16}
\end{figure*}

The evolution of the (\ej, $x$) characteristics of the main families of periodic orbits for Model 3 is given in Fig.~\ref{char3Dd16}. In contrast with Model 2, the x1 characteristic is stable also at the raising part of the curve beyond \ej $\approx -237000$. Simultaneously, already for \ej$\gtrapprox -232000$, there exists a stable 4/1 family, symmetric, with orbits of rhomboidal morphology. The families attributed to higher order resonances (7/1, 8/1) still exist, and even at a larger Jacobi constant interval than in Model 2. This time we find a stable part at the 7/1 characteristic, as well as a small stable part of 8/1. The morphology of the orbits of these families is similar to that of the corresponding families of the planar model (Fig.~\ref{pospir2d16}). Finally, we mention the presence of several 3/1 families, also with stable parts, at $-233000 \lessapprox$ \ej $\lessapprox -224000$.

\begin{figure}
\begin{center}
\includegraphics[width=8cm]{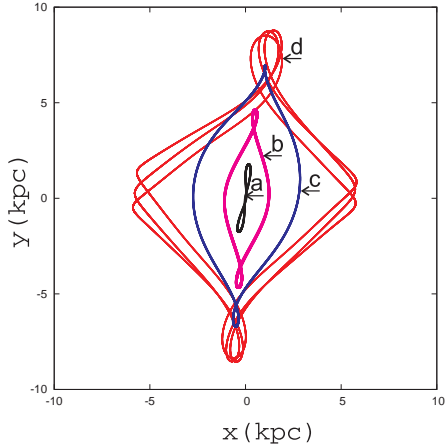}
\end{center}
\caption{Bar building orbits in Model 3. The two innermost orbits
belong to x1, while the two outermost to the 4/1 rhomboidal family.}
\label{bs3Dd}
\end{figure}

Looking first for bar supporting orbits, we have found that they belong to x1, 4/1 and even to 3/1 families. In Fig.~\ref{bs3Dd} we present stable bar supporting periodic orbits with loops, that can be considered as a skeleton of a bar structure. An arrow labeled ``a" points to the innermost orbit, which belongs to the x1  family and has  \ej = $-250000$. The morphological evolution of x1, also in this model, follows that of the propeller orbits described in \citet{kp05}. Orbit ``b", belongs also to the x1 family, has \ej = $-235000$, i.e. it is at the beginning of the raising part of the x1 characteristic. The third orbit of the sample, named ``c", is from the 4/1 rhomboidal family at \ej = $-230000$ and has asymmetric loops, with the lower loop being larger than the upper one. Finally the outermost periodic orbit, ``d", is an orbit of multiplicity 3, the characteristic of which is the segment, like an extension, at the local maximum of the 4/1 characteristic. This orbit has \ej=$-224000$.

\begin{figure}
\begin{center}
\includegraphics[width=8cm]{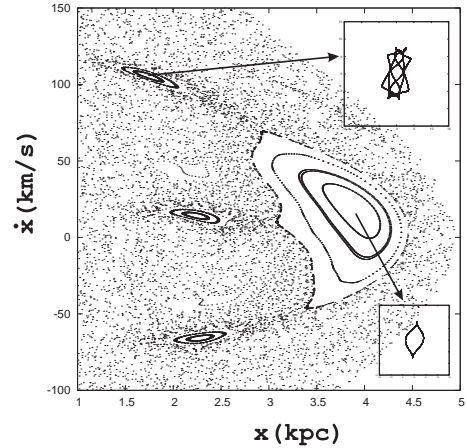}
\end{center}
\caption{Part of the surface of section of Model 3 for
\ej=$-228000$. The big island belongs to a stable orbit of the 4/1
family, while the three islands to a triple periodic orbit. The
morphology of the two orbits is given in boxes.} \label{sos3Dd16}
\end{figure}

Model 3 differs from the other models studied up to now, in that
there are always stable bar supporting orbits up to the 4/1
resonance region. Besides the stable simple periodic orbits the
surfaces of section are characterized by the presence of islands of
stable periodic orbits of higher multiplicity and sticky regions
around them. We give a typical example in Fig.~\ref{sos3Dd16}. It is
the (x,\.{x}) surface of section for \ej=$-228000$. The stable
periodic orbit at the center of the island of stability belongs to
the 4/1 family. Its shape is indicated in a box at the lower right
corner of the figure. Another stable periodic orbit at this
Jacobi constant is a periodic orbit of multiplicity three.
The three islands of stability, to the left of the 4/1 island,
belong to it and its shape is presented in a box at the upper left
corner of Fig.~\ref{sos3Dd16}. Note the darker areas, that appear as
bridges between the islands of the triple orbit and the island of
stability of the 4/1 orbit. As we can observe from the morphology of
these two periodic orbits both of them are bar supporting. Bar
supporting are also the sticky and other chaotic orbits that we
integrated for time equal to seven pattern rotations. We can say,
that the bar of Model 3 is built to a large extent in the classical
way, i.e. with regular orbits trapped around stable periodic orbits.
This bar has many common features with the NGC~1300 bar.


\section{Model 4}
We use Model 4 to present another mechanism, that we found to
support a spiral structure resembling that of NGC~1300. The
mechanism involves orbits around the Lagrangian points L$_4$ and
L$_5$. The efficiency of this mechanism in supporting spirals was
more evident in models with thick discs, especially in the range
20~\ksk $< \Omega_p <$23~\ksk\!. Model 4 is a thick disc model (see
PI) rotating with $\Omega_p =$22~\ksk\!.

\begin{figure*}
\begin{center}
\includegraphics[width=16cm]{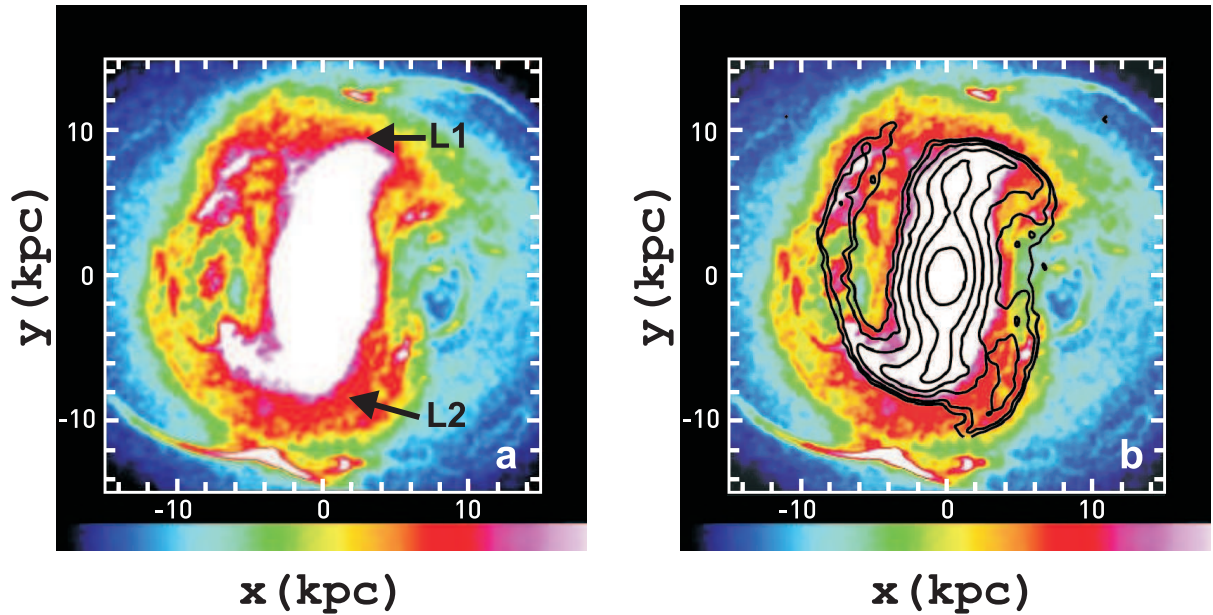}
\end{center}
\caption{The color-scaled density response of Model 4. It is a thick
disc model rotating with $\Omega_p =$22~\ksk\!. In (a) the location
of L$_1$ and L$_2$ are indicated with arrows. In (b) we overplot
characteristic isophotes of the K-band image of NGC~1300.}
\label{respo3Dd22}
\end{figure*}

\begin{figure}
\begin{center}
\includegraphics[width=6cm]{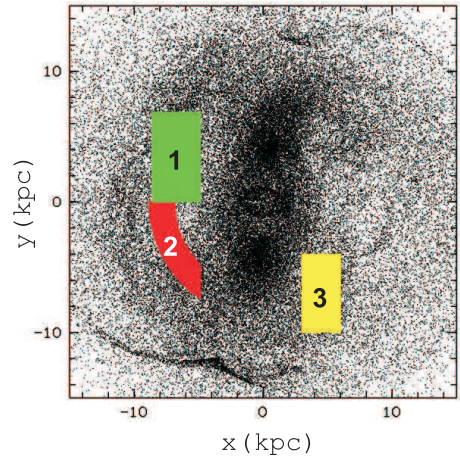}
\end{center}
\caption{The three regions (``1'', ``2'' and ``3'') at which we
examine the Jacobi constants of the particles participating
in the spiral structure of Model 4} \label{sele3Dd22}
\end{figure}

\begin{figure}
\begin{center}
\includegraphics[width=7cm]{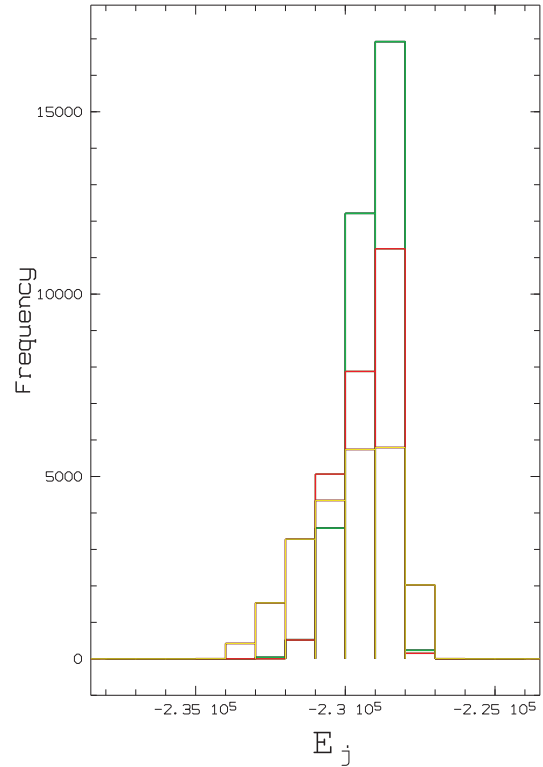}
\end{center}
\caption{Histogram giving the statistics of the
Jacobi constants at the three selected regions of the
spirals in Model 4. Green and red  correspond to the  regions ``1''
and ``2'' respectively in Fig.~\ref{sele3Dd22}, while the particles
of region ``3'' are represented by the yellow-grey bars.}
\label{histo3Dd22}
\end{figure}

\begin{figure}
\begin{center}
\includegraphics[width=8.2cm]{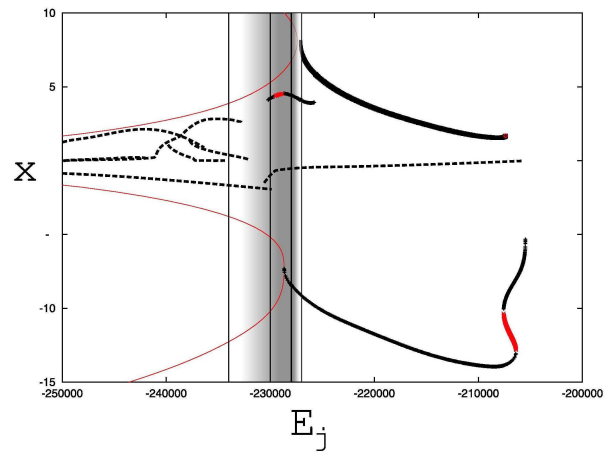}
\end{center}
\caption{The (\ej,$x$) diagram for the main families in Model 4.
Only the families drawn with heavy lines play a role in the
formation of the spirals of the model. The vertical lines indicate
the extent of the interval of Jacobi constants we study
(see text).} \label{char3Dd22}
\end{figure}

\begin{figure}
\begin{center}
\includegraphics[width=9cm]{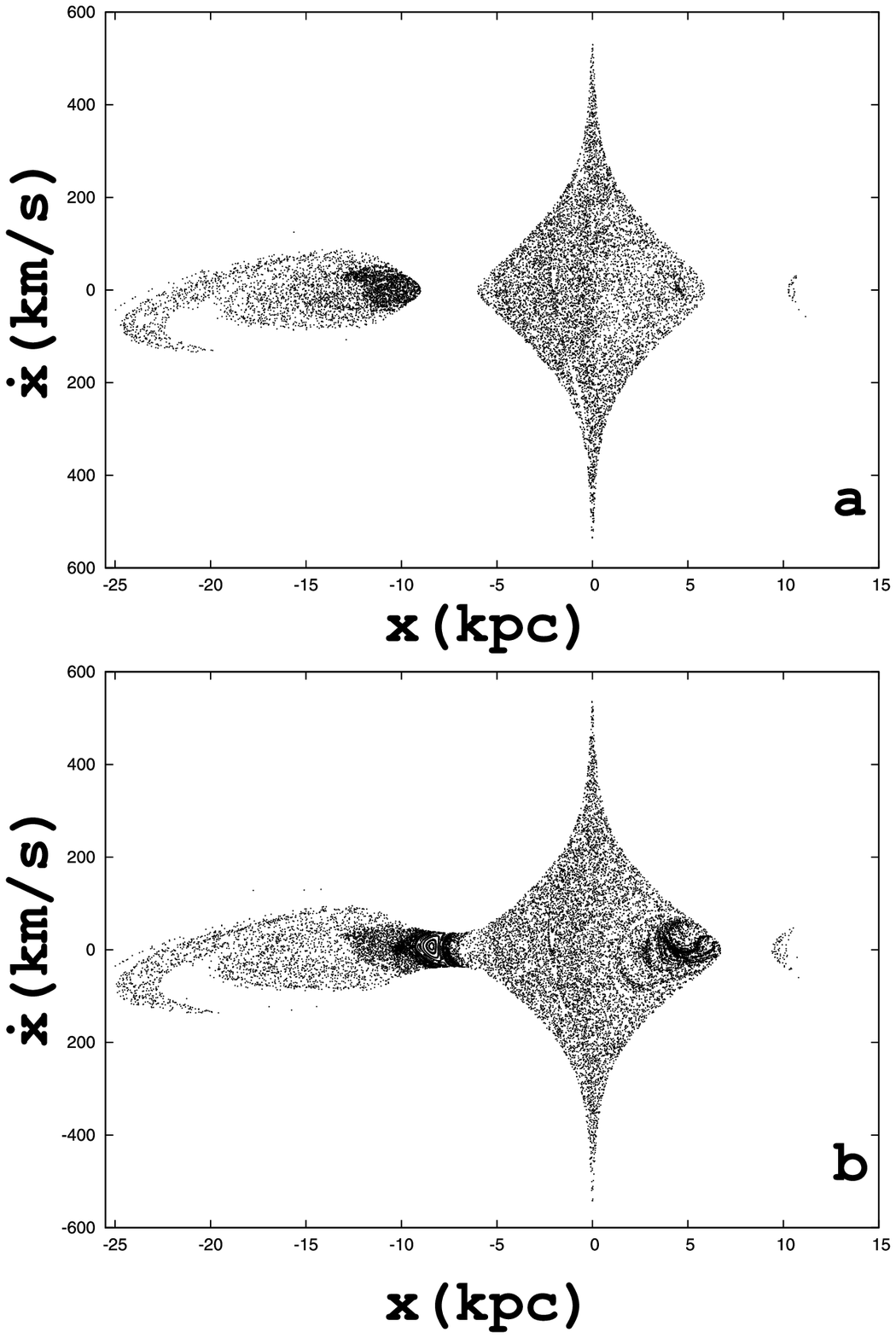}
\end{center}
\caption{Surfaces of section at  \ej=$-229000$ in (a) and for
\ej=$-228000$ in (b) in Model 4. At this Jacobi constant we
find most particles contributing to the spiral structure.}
\label{sos3Dd}
\end{figure}

The overall response of the model and the relation of it with the
K-band morphology of NGC~1300 is given in Fig.~\ref{respo3Dd22}. The
best reproduced feature of this model with respect to the rest of
the models we have examined until now is the beginning of the left
spiral, which emerges from the lower end of the bar. There is
clearly an increase of the local density of the model along the
direction, where the left spiral of the galaxy extends, as we can
see in Fig.~\ref{respo3Dd22}b. We also can observe that the right
spiral arm of the model breaks at about the same region as the
spiral of the galaxy and also that there is an increase of the local
density of the model close the lower end of the bar, at the right
side, in general agreement with the fragment of spiral arm we find
at the same region at the image of the galaxy. A feature of the
effective potential of this model worth pointing out is the
characteristic asymmetry of the Lagrangian points L$_1$ and L$_2$
with respect to the major axis of the bar. The $R_{CR}/R_b$
ratios we find for this model considering the L$_1$ and L$_2$
Lagrangian points, are 1.15 and 1.16 respectively. We have taken
three sets of particles from three dense regions of the model
spirals in order to find their Jacobi constants. This is
given in Fig.~\ref{sele3Dd22}, where these regions are
labeled with ``1'' (upper part of the left arm), ``2'' (lower part
of the right arm) and ``3'' (spiral segment at the right side,
close to the lower end of the bar). The statistics of the
Jacobi constants are given in Fig.~\ref{histo3Dd22}. Green
and red correspond to the the boxes labeled with ``1'' and
``2'' in Fig.~\ref{sele3Dd22} respectively, while the histogram of
region ``3'' is given in yellow-grey. The peak of the
Jacobi constants of the particles in all three regions is
at $-229000 <$ \ej $<-228000$, while we observe, that the
Jacobi constants of all particles we consider, have
$-234000 <$ \ej $<-227000$. So we focus our investigation to spiral
supporting orbits at these Jacobi constants. 
As regards
periodic orbits, we give in Fig.~\ref{char3Dd22} with heavy lines
the (\ej,$x$) characteristics of the families that are involved in
the enhancement of the spiral arms in Model 4. The left and right
vertical lines indicate the extent of the \ej interval that we have
to investigate as results from the statistics presented in
Fig.~\ref{histo3Dd22}, i.e. $-234000 <$ \ej $<-227000$. The two
lines between them show the part of the interval, where we have the
two highest bars of the histogram, i.e they are at \ej=$-230000$ and
$-228000$ respectively. The darkness of the grey shade reflects the
number of particles at the bins of the histogram in
Fig.~\ref{histo3Dd22}. Essentially the interval of the
Jacobi constants we study includes parts of the
characteristics of three families with stable parts. L$_4$ and L$_5$
are stable in this model. We study two characteristic surfaces of
section for two close by values of \ej. In Fig.~\ref{sos3Dd} we
present them for \ej=$-229000$ in (a) and for \ej=$-228000$ in (b).
In (a) we are located still before the local maximum of the ZVC at
negative $x$'s, close to L$_5$. The surface of section is
practically chaotic everywhere. However, we discern a darker region
at the right side of the left part of the surface of section. In (b)
the two regions, the central and the left one, have merged and a
family of stable banana like orbits, originated at L$_5$ appears
(invariant curves in the dense region at $x \approx
-10$~kpc). It is surrounded by a sticky region, while another dark,
sticky, region can be observed at the right part of the central
region. We followed the same procedure as in the previous models and
we integrated for time equal to 7 pattern rotations a large number
of initial conditions on both surfaces of section of
Fig.~\ref{sos3Dd}.

\begin{figure*}
\includegraphics[width=15cm]{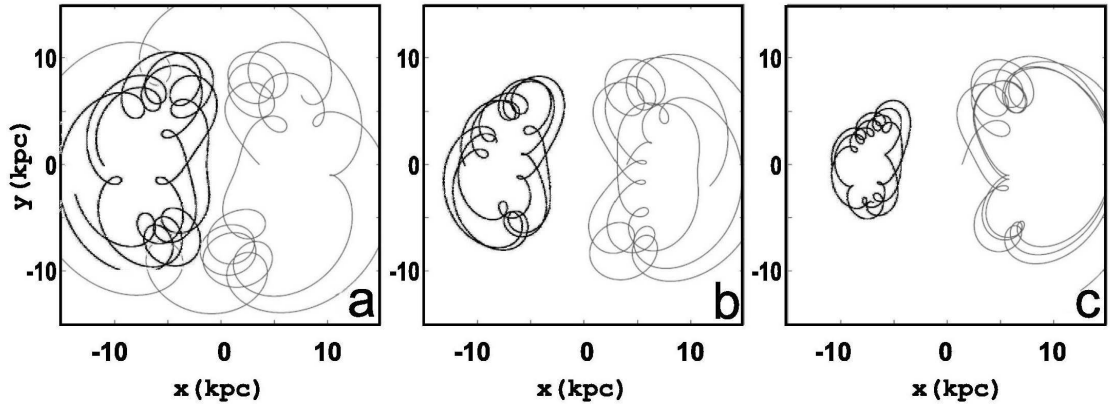}
\caption{Orbits at the sides of the bar in Model 4, with
\ej=$-229000$. With black we give orbits at the left and with grey
at the right side of the bar.} \label{orb3Dd_229a}
\end{figure*}

\begin{figure*}
\includegraphics[width=15cm]{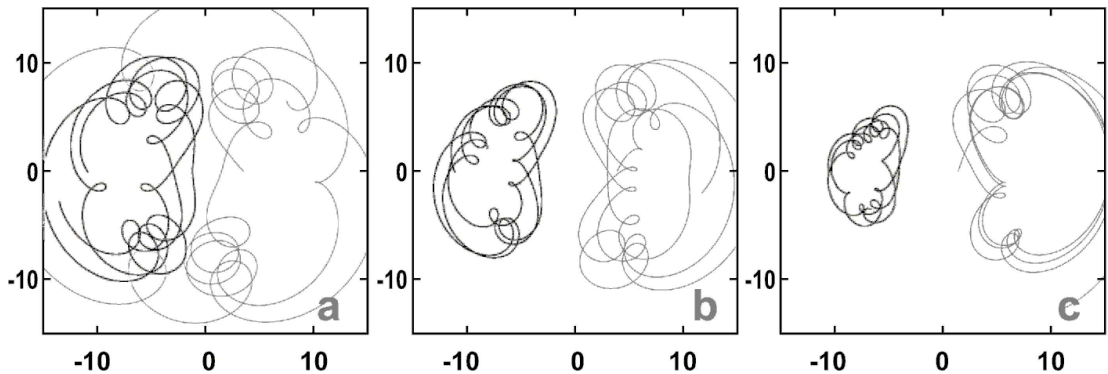}
\caption{Orbits at the sides of the bar in Model 4, with
\ej=$-228000$. Black orbits surround L$_5$, while grey L$_4$.}
\label{orb3Dd_228r}
\end{figure*}

In the subsequent figures we group the orbits of the
particles on the spirals as follows: Figure \ref{orb3Dd_229a}
presents characteristic orbits at the sides of the model bar. With
grey are the  trajectories at the right side. We observe the loops
at areas that correspond to the ends of the right side of the bar.
Simultaneously these orbits leave rather empty the region between
5$<x<$10~kpc. With black are given orbits at the left side of the
bar this time. Most of the loops of these orbits are along the
region of the left spiral arm of the model, while at their lower
left side they perform in most cases a single loop that brings them
closer to the bar. More efficiently acts this mechanism at
\ej=$-228000$. For the same reasons we explained in the case of
\ej=$-229000$, the orbits at Fig.~\ref{orb3Dd_228r},
which have \ej=$-228000$, contribute to the formation of the spiral features. With black we give orbits of various sizes surrounding L$_5$. The loops of the large size orbits enhance the lower and upper part of the left spiral arm, while the small sized contribute to the formation of the central part. With grey we give orbits surrounding L$_4$. 

Summarizing the response of Model 4 we can say, that density maxima
aligned along the location of the spiral arm features of NGC~1300,
are provided by the banana-like orbits. It is remarkable also that
the characteristic break of the right arm, just after it emerges
from the bar, and the presence of the spiral segment close to the
lower right end of the bar are also reproduced in an acceptable
degree by Model 4 (Fig.~\ref{respo3Dd22}). However, the drawback of
the mechanism is that one cannot get rid of the additional features
that are brought in the model from the same orbits that reinforce
the spirals. As we can see in Fig.~\ref{respo3Dd22}b, there is a lot
of additional structure that appears together with the spirals to
the left of the spiral arm. This structure is formed by the part of
the orbits that is far from the bar. It is a quite robust feature
and cannot be smoothed out even by increasing the dispersion in the
velocities of the initial conditions of response models with $20 <
\Omega_p < 24$~\ksk, always in the thick disc approximation.

We finally note that in Model 4 we have the clear evolution of two
spiral features that start with an overdense region almost on the
major axis of the bar, but away from it. The lower of the two starts
in Fig.~\ref{respo3Dd22} at about $(x,y)\approx(-2,-14)$ and goes
out of the frame at $(-15,-8)$. This feature, as well as its
symmetric with respect to the center of the galaxy at the upper part
of the model, are related with orbits associated with the outer
resonances, beyond corotation \citep{kc96}. However, it seems that
they do not have any correspondence in the near-infrared morphology
of NGC~1300.

\section{Other cases examined}
The class of models that is not represented in our analysis is the
one that assumed a spherical component representing the major part
of the bar (without the ansae) (see PI). This class of models did
not improve the resemblance between model and galaxy. However, in
PII we have seen that the model of this class especially with
$\Omega_p =21$~\ksk had a closer resemblance to the galaxy in its
very central part. The orbital analysis has shown that the inclusion
of an extended spherical component introduces in the system
circular, and nearly circular, x1 orbits at the major part of the
characteristic of this family. The characteristic remains close to
the $x=0$ axis at long \ej intervals. In the particular model the
presence of the circular x1 orbits secures a round central region of
radius $\approx$ 0.5~kpc. For larger Jacobi constants the
orbits of the x1 family become asymmetric with respect to the
x-axis. Their morphology becomes elliptical with one focus at the
origin of the axes of our system and the other at negative y. This
follows the tendency of the central region of the galaxy to become
elongated towards negative y. The results of all models presented
here do not alter qualitatively even if we chose different
projection parameters on the sky for the disc of NGC~1300. We find
similar dynamical mechanisms acting also for the projection
parameters (PA, IA)=106.6$^\circ, 42.2^\circ$), which are those one
finds by fitting an exponential disc to the regions outside the main
bar in the K-band image, as in \citet{gpp2004} in agreement with
\citet{ls02}. A few examples have been presented in \citet{pk09}.
The dynamical mechanisms are similar, the $\Omega_p$ numerical
values however, differ.

\section{Discussion and Conclusions}
Following the framework established in PI, we performed extensive
modeling of the stellar component of NGC~1300 in three classes of
models reflecting different geometries. As we have seen in PII, we
obtained acceptable solutions for the bar or the spirals not only
for more than one pattern speed, but for different geometries as
well. Anyhow, the main parameter that determines the morphology of
our response models is $\Omega_p$. The orbital analysis of the
present paper, shed light on the various ``promising'' models and
provided feedback, that could be used for the determination of the
optimum solution. However, the most interesting result was not the
determination of the optimum parameters for a ``best'' NGC~1300
model, but the dynamical mechanisms themselves, which are
leading to the desired morphology. Even models to which our analysis
was not pointing to as best matching the NGC~1300 morphology, have a
special interest, since their parameters are equally realistic for a
barred-spiral system in general as those best applying to the
specific galaxy.


Summarizing the four models we have chosen to present, we found that there have been two values of $\Omega_p$ for which our models were giving the best matching morphologies to NGC~1300 in the three classes of models we considered, namely $\Omega_p$=16~\ksk and 22~\ksk\!. The new dynamical mechanisms that we found deserving special attention and extensive presentation, were the following:
\begin{itemize}
 \item In a 2D model rotating with $\Omega_p$=22~\ksk\!, a strong bar with ansae is supported practically \textit{only by chaotic orbits}. This bar has similar morphology with the bar of NGC~1300. Decisive role in the support of such a structure by chaotic orbits plays the presence on the effective potential of multiple Lagrangian points roughly along the major axis of the bar. We can understand this by considering an $\Omega(r)$  curve, which is not monotonically decreasing, but has a wavy character. Then it is possible to get the same $\Omega_p$ at two (or more) points along the $\Omega(r)$ curve. This is the case of the bar in our Model 1. It is an alternative mechanism for the support of the ansae morphology based only on chaotic orbits. Probably it can be applied to many bars showing this feature.
 \item In a 2D model rotating with $\Omega_p$=16~\ksk\!, our Model 2, the morphology of the spiral arms of NGC~1300 can be reproduced by chaotic orbits at \ej's between the 4/1 resonance and corotation. In our case the \ej values of the orbits that support the spirals are well inside corotation. The spirals are supported by the ``bouncing" of the chaotic orbits on the ``walls" of the effective isopotential drawn with a heavy line in Fig.~\ref{effpot2d16}.
 \item In a thick disc model rotating with $\Omega_p$=22~\ksk\!\!, we found a mechanism supporting a spiral structure similar to that of NGC~1300 with orbits around the Lagrangian points L$_4$ and L$_5$ (our Model 4). These orbits are either regular orbits trapped around stable periodic orbits at the region, or sticky chaotic orbits, or even chaotic trajectories from a pool of a chaotic sea. All of them contribute to the support of the spiral arms. The possibility that a set of banana-like orbits can support a spiral morphology has been indicated by \citet{p06} and \citet{gco09} \citep[see also][]{cp06}.
\end{itemize}
We do not include in our enumeration of the new mechanisms the formation of a typical ansae bar by regular orbits as in our Model 2.

By ``chaotic orbits'' we actually mean trajectories, that occur after integrating for a given time initial conditions, which we select out of a common chaotic sea, i.e. essentially we examine fragments of a single chaotic orbit. If at a certain \ej a percentage of these trajectories supports a given structure (a bar or a spiral) it practically means, that the orbit that fills the chaotic sea supports this structure during the same percentage of its, long, integration time. Our result, that surfaces of section without any discernible island of stability provide support to a specific morphological feature, means that the chaotic sea is structured in such a way, that allows this dynamical behaviour. Despite the fact, that there is no obvious structure in a chaotic sea, it is known that
the manifolds associated with the unstable periodic orbits existing on the surface of section do structure the chaotic sea \citep[e.g.][]{gco02}. In our models the spirals that we discuss as reproducing the spiral arms of NGC~1300 are \textit{inside corotation}. However, the general relation between manifolds and resulting spiral structure is expected to be similar to the one described by \citet{tev08}, \citet{tev08} and \citet{tkec09} for spirals related with the family of unstable periodic orbits originated at the unstable Lagrangian points L$_1$ and L$_2$ \textit{at corotation}. We do not point to a specific family of unstable orbits, the existence of which determines that of the spirals in Model 2 or the ansae type bar in Model 1. It is the collective dynamical behaviour in chaotic seas within a specific \ej range. The dynamical mechanisms that shape a structure are in general different. In Model 2 it is the restriction of the chaotic orbits by a characteristic isocontour of the effective potential and the loops of the orbits as they ``bounce'' on it. On the other hand in Model 1 it is the trapping of the orbits in the central area and the presence of the double L$_1$ points that shape the morphology of the bar. The structures that are supported by chaotic orbits are dictated to a large extent by the shape of the isocontours of the effective potential.
The fact that star clusters in the spiral arms of NGC~1300 are not well aligned as in other grand-design spiral galaxies (e.g. NGC~2997) \citep{gd08} is in agreement with a chaotic character of the orbits of the stars that build them.

It has been proven quite difficult to point to a best NGC~1300 model. The assumption of different pattern speeds for the bar and the spirals seems less probable, due to the relative locations of the two components. This is not concluded from the apparent connection of the spiral arms and the bar, but mainly from the fact that we do not find two different, separated zones, where bar and spirals rotate without interacting. A bar rotating faster than the spirals will reach the spiral arms at a certain fraction of the spiral period and then there will be an interference between the two components. In such a case the observed morphology would be just a snapshot in a transient morphology of the barred-spiral system. An indication against the transient spiral hypothesis is that the
spiral arms exist in K-band and they are even sharp features as we mentioned in the introduction. If we have to deal with a more or less stationary morphology, the regions where bar and spirals rotate should be rather separated. As an example we mention the NGC~3359 model by \citet{pkgb09}, where in a two pattern speeds model the orbits of the spiral component change their topology in such a way as to allow the bar to rotate within a certain radius leaving the spirals outside it (their fig.~20).

From the large number of response models we presented in PII it was evident that it was more difficult to find a response spiral matching the spiral arms of NGC~1300, than to model the bar. The reason is that at a certain range of $\Omega_p$ values, in all three general classes of models we studied, the length of the bar is close to the length of the bar of the galaxy. From the two mechanisms that we found supporting the galaxy's spiral, the one in Model 2 (chaotic orbits between 4/1 and corotation) has more advantages than that of Model 4 (regular and chaotic orbits around L$_4$ and L$_5$). Model 2 follows closer the pitch angle of the left arm of NGC~1300, while the spiral fragment at the right of the lower end of the bar is more conspicuous in Model 2 than in Model 4. The chaotic spirals of Model 2 with planar geometry can be combined with the bar of the thick disc Model 3 in a model with a single pattern speed $\Omega_p$=16~\ksk\!. On the other hand the left spiral of Model 4 is combined with additional features (Fig.~\ref{respo3Dd22}, which we cannot ignore in the comparison with the K-band image of the galaxy. There are regions of density maxima to the left of the left arm, that are formed due to the same orbits that give the density maxima along the response density maxima corresponding to the left arm of NGC~1300. These superfluous features could not be eliminated by increasing the velocity dispersion of the initial conditions. By combining Model 1 with Model 4 in a single model rotating with $\Omega_p$=22~\ksk\!, we have a 2D bar combined with a spiral in a thick disc beyond the bar. Thus, taking into account the advantages and disadvantages of all models we have found having a good relation with the NGC~1300 morphology we want to model, we conclude that Model 3 for the bar and Model 2 for the \textit{stellar} spirals, both rotating with $\Omega_p$=16~\ksk\!, give the best representation of the corresponding structures of NGC~1300. We underline that the two preferable $\Omega_p$ values (16 and 22~\ksk) are not chosen among equally good models with nearby pattern speeds, but it is a result of the extensive modeling presented in PII, that we have best responses for these two values.

A comparison of our results with the modeling of NGC~1300 by other authors has been done in PII. We want here to add that our Model 1 gives Lagrangian points (Fig.~\ref{effpot2d22}) that are close to the corotation radius of the ``fast"  model by \cite{lk96}, 104$\arcsec$, (L$_1$, L$_2$), as well as with the corotation radius of \citet{e89}, $\approx$83~kpc, (L$_{1S}$, L$_4$, L$_5$). As we explained there is no need for a two pattern speeds assumption for having multiple Lagrangian points in a model.

Below we enumerate our conclusions:
\begin{enumerate}
 \item We could reproduce structures similar to those of the bar or the spiral of NGC~1300 by more than one model. All models presented in the paper have some nice agreement with the morphology of the galaxy. The comparison of these models with the desired barred-spiral morphology, as well as the comparison among themselves, indicates as best a configuration with a single pattern speed, where the whole barred-spiral structure extends inside corotation. The bar apparently is composed by regular orbits (Model 3), while the \textit{stellar} spirals are composed by chaotic orbits (Model 2).
 \item We find an ansae-type bar morphology that can be constructed completely out of chaotic orbits (Model 1). This requires the presence of multiple unstable Lagrangian points roughly along the major axis of the bar in the effective potential (Fig.~\ref{effpot2d22}). This is an alternative way of building ansae-type bars.
 \item Chaotic orbits with Jacobi constants between 4/1 and corotation can build a spiral. The dynamical mechanism, described in  our Model 2, is based on the loops these orbits perform at their apocentra, as they move in a region defined by the isocontour of the effective potential that separates the regions inside and outside corotation.
 \item Another dynamical mechanism that can support a spiral is based on banana-like orbits. These orbits are regular as well as chaotic (Model 4).
 \item The recently discussed spirals, that are associated with the stable and unstable manifolds of the unstable Lagrangian points L$_1$ and L$_2$, along the major axis of the bar, do not play any important role in our models. We find these spirals in some cases, as the one of Model 1. They emerge from the ends of the bar and fade out after completing an azimuthal distance of about $\pi/2$. However, they are never in agreement with the NGC~1300 spirals.
 \item The morphology of the bars and spirals that are supported by chaotic orbits in our models are determined to a large extent by the shape of the isocontours of the effective potential (Figs.~\ref{effpot2d22} and \ref{effpot2d16}).
\end{enumerate}


\section*{Acknowledgments}
We thank Prof. G.~Contopoulos for fruitful discussions. We also thank an anonymous referee for many constructive comments. P.A.P
thanks ESO for a two-months stay in Garching as visitor, where
part of this work has been completed. All image processing was done with the ESO-MIDAS system. This work has been partially supported by the Research Committee of the Academy of Athens through the project 200/739.

\label{lastpage}

\end{document}